# Developments in the negative-*U* modelling of the cuprate HTSC systems

John A. Wilson.

H.H. Wills Physics Laboratory,

University of Bristol,

Tyndall Avenue,

Bristol BS8 1TL. U.K.

e-mail; john.a.wilson@phy.bris.ac.uk

**Abstract**

This paper emphasizes once more the many strands which go into creating the unique and complex nature of the mixed-valent HTSC cuprates, above $T_c$ as below. Clearly it is not sensible to look in isolation to the lattice, magnetic or electronic aspects of the situation. Circumstances are too tightly coupled and interdependent. What one is in a position to do is to ascribe some pre-eminence in these matters. The line taken in this paper, as in its predecessors, is that charge constitutes the prime mover, but this taken within a chemical context, relating to bonding effects in the copper oxides and to a crucial shell-filling negative-*U* term of large magnitude. What dictates the uniqueness of copper in this is its particular position within the periodic table at the end of the 3*d* series. This provides access to (i) three stable valencies, (ii) to the limited metallicity of tight-binding, mixed-valent $YBa_2Cu_3O_7$, etc., (iii) to the *p*/*d* hybridized limitation of magnetic behaviour, and (iv) above all to the shell closure effects, with their dramatic energy adjustments of the relevant double-loading states when sited in high-valent local environment. This behaviour rests on the inhomogeneous electronic and structural conditions existing within these tight-binding mixed-valent systems. Normally such systems show strong positive-*U* behaviour and are magnetic. Here the shell-filling effects negate this *U* term to leave a net negative $U_{eff}$ of –3 eV per pair. That positions the bosonic electron pairs degenerate with $E_F$. RVB spin coupling within the stripe phase geometry set up by the doped charge removes much spin-flip pair breaking, while the adopted 2D crystal structure with its saddle-point dominated F.S. provides the ideal *k*-space geometry from which to secure the negative-*U* driven pair formation. The Jahn-Teller effect associated with the $d^9$ electron count is crucial in upholding the two-subsystem nature of the HTSC materials.

This paper looks at the support for this scenario that can be extracted from the recent work of Corson, Li, Mook, Valla, Norman, Varma, Gyorffy and many others. To the author it remains a mystery why other researchers have not examined the potential of this route for understanding the unique characteristics of the HTSC cuprates.

**Contents**

## 1. Brief resumé of the chemical negative-U approach to HTSC.

There is still surprisingly little integration to be found between the great variety of theoretical approaches towards the subject of cuprate HTSC. The bank of experimental data is today of course vast and complex, and pursuing the formalities of some highly restricted 'toy model' plainly is not what at this point is required to resolve the problem. Indeed I would claim that virtually from the start it has been apparent that the main impediment to solving the HTSC conundrum has been a general reluctance to take just one step beyond the bounds of traditional physics to address precisely how cuprates, and these cuprates in particular, differ from all other materials. The fact that they most clearly do should stand as primary input to any theoretical modelling. What is so unique about copper, about its position within the periodic table and about what this entails for the properties of its compounds? Why is the HTSC phenomenon so confined to copper oxides, and then solely to mixed-valent, layered, square-planar coordinated ones? Not nickel, not silver, not titanium, not ruthenium, just copper. I do not wish here to go back over these matters in detail; the literature already bears a heavy load [1]. There exists however one item of 'chemistry' that needs underlining; namely what arises within any compound with regard to intersublattice bonding/antibonding behaviour. Whether largely covalent or partly ionic such interaction can generate band shifts of many eV away from the energies of the parent atomic states involved. Under tight-binding conditions such movement continues to be evident at the local level within interorbital and intersite charge fluctuations. No effects of this type are more marked than when some charge transfer closes out a quantum shell, terminating thereby all bonding/antibonding interaction for the states and sites concerned. Especially for 3*d* systems the antibonding states then are released to fall by several eV, particularly when in high valent local environment, to become incorporated both energetically and spatially into a spin zero, outer core condition. The specific shell closure fluctuation effected by pairs of electrons passing onto a nominally trivalent cuprate coordination site (8 electrons or 2 holes) is, it is claimed, the crucial factor consistently being disregarded when addressing the HTSC problem [1]. Being fluctuations such changes involve a somewhat smaller energy change than otherwise. Nonetheless in the present case a very sizeable effect still results because the fluctuations in question occur into formally trivalent coordination units. These electronic, 'chemical', negative-*U* fluctuations are able there to yield effects much greater than those secured under the alternatively advocated phonon, lattice bipolaron, and magnon-type spin-fluctuation routes to HTSC.

From the beginning I have drawn attention to the fact that these mixed-valent systems are inhomogeneous, structurally and electronically [1a], and in ways which are of direct import to the appearance of HTSC. It took only the incommensurate neutron diffraction data of [2] to prompt a conversion of the dopant randomized figure 4 in [1a] to the charge and spin structured form appearing in figure B1, etc. of [1d], following our comparable work on the discommensurate CDW arrays discovered in layered 2H-$TaSe_2$, $1T_2$-$TaS_2$, etc. [3]. While this ‘stripe phase’ structuring, and the Jahn-Teller effect to which in the cuprates it is coupled, are of considerable significance in controlling many of the detailed aspects of the cuprate data [4], it however remains the still-local valence inhomogeneity which exists within these mixed systems of two Mott insulators (eg. $La_2CuO_4$ + $LaCuO_3$) that predicates why in HTSC we are dealing with oxides, and with 3*d* oxides in particular, not 4*d* or 5*d*. For the HTSC cuprate materials the way that their square-planar crystal chemistry controls the band structure – and hence the geometric form to the Fermi surface and its local dispersion characteristics – is clearly too a vital matter and one again of detailed record [1f]. However this physics already constitutes a secondary stage within the genesis of HTSC. The primary input is without doubt the matter of the Cu/O bonding/antibonding interaction, and in particular the p/d shell closure attained under the mixed-valent, intersubsystem, double-loading fluctuation between antibonding $d_{x2-y2}$ states that is represented in [1a] by

$$^{9}Cu_{II}^{0} + {}^{9}Cu_{II}^{0} + {}^{8}Cu_{III}^{0} \longrightarrow {}^{8}Cu_{II}^{1+} + {}^{8}Cu_{II}^{1+} + {}^{10}Cu_{III}^{2-} .$$

The symbols disclose here first the site formal valence (Roman subscript), secondly the instantaneous site electron count, and thirdly any deviation in site charge away from the non-fluctuational condition.

It was noted in [1a] that bonding/antibonding effects have a certain capacity to be of wider relevance to the generation of strikingly raised superconducting $T_c$’s; witness the ubiquitous superconductivity of the homopolar bonded, high pressure, semimetallic, non-close-packed forms of many elements from groups IVB-VIB, like Si, P and S [5]. When one turns to look at homopolar bonding within a compound environment, we have, since the beginning of HTSC, additionally seen very remarkable superconductivity emerge in such diverse homopolar bonded materials as $Rb_3C_{60}$ (carbon balls) [6], $Li_x$HfNCl.yTHF (bilayer hafnium nets) [7] and most recently $MgB_2$ (monolayer boron nets) [8]. My introductory paper of 1987 [1a] actually remarked on the potential in this regard of the semimetallic layered diborides (§7.1). One might further point to the $Mo_6$ cluster Chevrel phase superconductors such as $PbMo_6S_8$, and even to tetrahedrally cross-linked A15-type $Nb_3Ge$, immediate forerunner to HTSC.

With regard to $T_c$ all the above systems present an isotope effect of some form, even the cuprates. This observation does not however imply that electron-phonon coupling must be assigned prime responsibility for the superconductivity. Strong electronic effects inevitably introduce strong lattice coupling. In the cuprates such coupling may be observed directly by employing such local structural probes as PDF and EXAFS [9]. The HTSC data reveal the Jahn-Teller distortion associated with the $d^9$ condition to be especially marked [4]. It is not in doubt, either, that strong lattice polaronic effects are in play in the cuprates, for which carriers in the underdoped regime stand not too far removed from Mott-Anderson localization under the strong correlation and valence disorder existing there. It nonetheless is plainly necessary to advance beyond electron-phonon [10] and polaronic [11] claims to being prime mover if one is to encompass fully all the accumulated data within the HTSC cuprate domain. Witness for example the gyrations displayed versus doping content of the isotope effect [12], the temperature dependencies of the electronic and thermal Hall effect results [13], and the unusual Raman spectra both electronic and phononic [14].

Previously with the Bucky Ball materials [15] and now too with $MgB_2$ [16] a great many people still turn toward standard electron-phonon coupling as primary source of the new superconductivity. However a $T_c$ in $MgB_2$ of 40 K, and one within recently produced *p*-type injection-doped $C_{60}$ of 55 K [17] must signal that there is stronger mediation involved here than traditional retarded electron-phonon coupling [18]. It will be most interesting to discover if by de Haas-van Alphen experimentation a Fermi surface is extractable for $MgB_2$, or whether once again electronic and structural fluctuation effects introduce too great a scattering. Correspondingly it would be highly illuminating to have a fresh attempt made to obtain dHvA signals from the perovskite bismuthate system $(Ba/Rb)BiO_3$, superconducting to 35 K [19]. From my own point of view this latter mixed-valent system stands closest to duplicating what I am claiming for the HTSC cuprates. In the bismuthates, however, with the phenomenon there being based upon *s*-shell closure fluctuations [1b], and with these being less local, $T_c$ emerges as less spectacular.

It long has been evident that the Fermi surface in the HTSC cuprates even at substantial carrier contents (i.e. screening) is highly perturbed by the various fluctuations and scatterings occurring there. Not only is the electrical resistivity in optimally doped samples linear in $T$ over all $T$ to very high $r$, but the Seebeck coefficient well in advance of the superconducting onset is increasing and large in magnitude [20], as too is the thermal resistance [21]. It is clear that much of the pseudogapping responsible, being strongly in evidence also within all types of optical data [22], is of a dynamic origin. Note that neither the specific heat [23] nor the static magnetic susceptibility [24] are particularly enhanced when one considers how close in the materials are to the Mott-Anderson transition. The current carriers below and above the $d_{x2-y2}$ band half-filling point of $d^9$, from their universally inverted *p*- and *n*-type characters respectively, afford further manifestation of the strong DOS pseudogapping and of just how non-standard the HTSC cuprate condition is. The electronic virtually magnetic field independent form to this most abnormal 'normal state', and to the superconductivity which develops from it, is very evident from the low temperature fall-off displayed below and indeed somewhat above $T_c$ in the nmr spin-lattice relaxation rate [25], in addition to the thermoelectric data [26,20]. Once below $T_c$ one should register too the rapid rise in electronic mean free path in evidence from the thermal Hall conductivity results [13]. The unusual $T^4$ temperature dependence exhibited above $T_c$ by the magneto-resistance [27;1d appendix A2.8] is a further oddity of similar issue.

The detailed configuration within $k$-space of this very strong scattering above $T_c$ and the attendant pseudogapping steadily has become much clearer through the high resolution probing now achievable both in neutron [28-39] and ARPES [40-58] experiments. The neutron work in particular is informative about those changes in spin state condition that are necessary in advance of the appearance of HTSC, in addition those which reflect it. Below $T_c$ a better view has been gained likewise of the dominantly $d_{x2-y2}$ form to the fermionic pairing order parameter current here.

## 2. Synopsis of some important developments in recent work other than neutron and ARPES bearing on the HTSC mechanism proposed.

Due to the advanced state of *p*/*d* hybridization encountered at the termination of the 3*d* series, most Cu systems on cooling progress through a rather special set of charge and magnetic circumstances (in addition to those more directly related to superconductivity itself). In fact charge delocalization proceeds here invariably through reduction in the *p*/*d* 'charge transfer gap' in advance of the Mott-Hubbard gap [59]. This delocalization encourages normal superexchange antiferromagnetic coupling to give way to spin-singlet dimer formation, as seen in ($^{9}Cu_{II}^{0}$) 1D $Sr_2CuO_3$ [60] or 2D $SrCu(BO_3)_2$ [61]. In the case of the valence-substituted HTSC cuprates, the low temperature drop-off in magnetic susceptibility has in [1d-f] been ascribed to RVB square-plaquet formation, within a charge-segregating stripe phase format. The resulting spin gap possesses then the same $d_{x2-y2}$ symmetry as the superconducting gap displays below $T_c$, although now upon increased underdoping (i.e. localization) this gap initially grows larger not smaller. That continues until RVB ultimately is replaced by SRO AF as the collapsing spin gap becomes surpassed by the rapidly growing Mott-Anderson charge gapping. This is apparent in the low temperature resistivity [62] and the Knight shift data [63]. By that point superconductivity is fully suppressed. Such modifications to the gapping status of the quasiparticle DOS may also be traced through tunnelling [64, 65] and IR measurements [66, 62], as well as in the photoemission [40-58] and neutron [28-39] work.

The spin pseudogap condition has been shown in [25b, 65] to be virtually unaffected by a strong magnetic field until quite close in to $T_c$. It is clear that the very extensive pseudogapping is not to be attributed simply to significant 'local pair' formation arising far above $T_c$, with those pairs then awaiting (in a fashion analogous to Fe spin moments) some low temperature onset of spatial coherence – in the present case of a uniform superconductive phase angle. In the advocated inter-subsystem pair-fluctuation scenario, appreciable long-term pair formation (as $^{10}Cu_{III}^{2-}$) only becomes possible as the electrons in the surrounding 'majority' fermion subsystem ($^{9}Cu_{II}^{0}$) are driven to undergo pairing through resonant pair exchange with the higher-valent subsystem. This interdependence means that extended-term, system-wide, delocalized pairing is withheld during cooling until only ten degrees or so in advance of $T_c$ itself. Nonetheless this constitutes a remarkably wide precursor superconductive behaviour, clearly in evidence in quantities like the diamagnetism [67], the paraconductivity [68] and the specific heat [23]. There are indications of the above developments too from the decrease in the current noise signal [69] and from modification to the rare-earth-based 4*f* crystal-field excitations in for example $HoBa_2Cu_4O_8$ [70]. The latter changes are the result of 4*f* coupling into the Cu 3*d* spin field and are recorded by means of inelastic neutron spin scattering. Interestingly these 4*f* crystal-field experiments detect too a very sizeable copper isotope effect which is observed to onset at much higher temperatures. A complemetary oxygen isotope effect already had been reported there from μSR [71], magnetization and Raman experiments [12]. Such behaviour may be taken to reflect the changes in bonding introduced with RVB and stripe phase development and the strong accompanying local Jahn-Teller distortions [9]. It is this way I believe that one should accommodate also the expansivity measurements reported recently by Meingast *et al* [72]. The latter workers have not considered the structural and magnetic changes associated with stripe phase formation, and instead attribute the observed anomalies in their entirety to superconductive fluctuations of an anisotropic 3D-*XY* Ising-like form. Superconducting fluctuations are conceived there to extend up to $2T_c$ for underdoped material, a consequence of Meingast *et al* regrettably adopting a standard homogeneous local pair viewpoint.

With progressive charge overdoping the stripes quickly become highly compacted and the RVB plaquets are driven out of existence, as time correlations of the spins and spin gapping fade away under the rapidly growing metallization. In consequence the low temperature static susceptibility in fact rises somewhat between $p \sim 0.16$ and 0.30 [73]. Results obtained from high resolution STM work [74] and also from electronic specific heat studies [23] actually indicate the spin pseudogap to drop below the superconducting gap by $p \sim 0.16$ (optimal doping w.r.t. $T_c$) on its way to zero at $p_c \approx 0.19$. One has to note, though, that even at $p = 0.19$ the number of condensed carriers does *not* reach the number of hole (let alone electron) carriers implicit in the given stoichiometry. Specific heat results, when appropriately analysed through integration of the electronic entropy [23], disclose furthermore a superconducting condensation energy (effective per mole of material) that is decreasing immediately and sharply from its peak value attained as $\Delta_{sp}$ became zero. This deduced fall off in $\Delta E_s$ develops at a far faster rate than μSR data would indicate the superconducting pair count itself to be falling [75]. It has been reported that upon yet further overdoping $n_s$ returns back sharply towards zero [76], although more recent μSR work [77] using well-formed LSCO did not find this in advance of $x = 0.24$. Rather it is the magnitude of the superconducting gap which is decreasing steeply. It is apparent that under the accelerating hybridization and the ensuing loss in differentiation between the two valence subsystems the chemical negative-*U* route to HTSC is losing its efficacy to promote pair binding.

It could from the above appear that in *under*doped systems there exists a latent capacity for $T_c$ to be considerably larger – the negative-*U* potential of the higher valent subsystem is there. However so too is the unquenched magnetism within the lower valent subsystem. Unless and until the magnetic tendencies within the latter are strongly constrained the resulting $T_c$ for the whole system remains much restricted. As was indicated in [1c] the role of the counter-ions in advancing $T_c$ is to facilitate this spin quenching at a lower carrier level via a broad band hybridization into the crucial Cu-O states near $E_F$: hence the favorable resort made to the 6*s*/6*p* elements Hg, Tl and Bi. The large size of these ions additionally tends to favour 2D structuring.

What is the evidence for the negative-*U* centres trying to seed superconductivity above the realized $T_c$ values but being denied by the residual magnetism? Experimentally it is well-established that anything which promotes magnetism by means of carrier localization or hybridization greatly reduces the $T_c$ attained: witness for example the striking consequences of Zn substitution for basal Cu or of Pr substitution for Y in YBCO etc. [78, 79, 1d,f]. One similarly might point to the consequences of the development of ordering in system microstructure closely associated with the $p = {}^{1}/_{8}$ count [4]. μSR [80], NMR [81] and NQR [82] experiments all reveal that in the vicinity of the latter carrier count there occurs a certain resurgence of unpaired magnetic moment formation at low temperature. The same conclusion is to be reached from neutron diffraction work on LSCO undertaken in a field of 10 tesla [83], where specifically for $x = 0.12$ $T_c$ becomes suppressed to 12 K and this in conjunction with increase in the intensity of the incommensurate spin scattering. It would appear here that individual vortex lines in the internal field bring about the elimination of superconducting behaviour over quite sizeable areas: $l$ is particularly large in HTSC materials (*i.e.* $H_{c1}$ is very small). Throughout the underdoped regime even a moderate field rapidly augments the Swiss cheese aspect to the superconductivity. The intrinsically micro-granular nature of HTSC materials becomes further evident in the results obtained by Loram, Tallon, Nachumi and coworkers when examining the electronic specific heat for $Tl_2Ba_2CuO_{6+\delta}$ in a magnetic field [84], and again when examining Zn-doped LSCO and YBCO by μSR [85]. Effective local magnetic organization is directly in evidence in the μSR experiment. At this point one may recall that magnetic pseudogapping has been demonstrated by means of STM spatial scanning [86] to persist within individual magnetic flux vortices below $T_c$. Inside a vortex, with the superconducting gap there now eliminated, the RVB gap of the low-valent subsystem is clearly to be seen.

What evidence next is there for the fast dynamics fundamental to the current two-subsystem modelling? Since the pair coherence length $x$ is so very small in HTSC materials, and since too the superfluid density $n_s$ is rather low, superconductive dissolution is going to be much more prone to phase fluctuations of the order parameter than in standard superconductors where Cooper pair unbinding is dominant. For HTSC materials the phase coherence times, $t_\phi$, are likely to be set to electronic scattering frequencies, not phonon frequencies. Indeed a simultaneous fluctuation-induced magnetoconductivity/paraconductivity/diamagnetism analysis of the Aslamazov-Larkin type [87] yielded a phase coherence time near $T_c$ of only $2\text{x}10^{-16}$s for untwinned high quality $YBCO_7$. This very short $t_\phi$ will be taken to relate to the majority divalent subsystem. There is reason to believe though, as we shall see, that within the key minority negative-*U* subsystem the pair lifetime near $T_c$ can stand up to ten orders of magnitude longer than this, while intersubsystem quasiparticle effects are operative on a sub-picasecond timescale. In this case in order to be able to investigate what is underway in the temperature range around $T_c$ one is drawn to consider a probe with a frequency ~ 500 GHz. Such submillimetre experimentation is rather hard to accomplish, whether optically or electronically. Corson *et al* [88] however have managed to obtain invaluable results in this waveband by means of time-domain transmission spectroscopy. They operated at a series of set freqencies from 100 to 600 GHz, and using films just 500 Å thick extracted directly both the real and imaginary parts to $s(\omega,T)$ for several *under*doped samples of BSCCO-2212 over a wide range of temperatures from 4 to 200K. Their intent was to determine the superconducting "phase stiffness" as given by the functional quantity $\bar{h}^2.n_s/m^* \equiv k_B T_\theta$ . $T_\theta(w,T)$ defines the characteristic 'stiffness temperature' (for the chosen composition). In the superconducting state $s(w,T)$ is given simply by $i.s_Q.(k_B T_\theta / \bar{h}\omega)$, where $s_Q \equiv e^2/\bar{h}d$ with $d$ the interplanar spacing in the quasi-2D superconductor. For such a situation we are directed to the treatment introduced by Kosterlitz, Thouless and Berezinskii [89] of pair correlation dissolution in terms of a defect vortex pair-binding (clockwise vortex with anticlockwise in phase angle terms), and of their subsequent thermal unbinding. If things really do proceed in this manner then one expects to observe a linear relation $T_\theta(w,T) = {}^{8}/_{\pi}.T'$ defining the crossover temperature $T'$ at which $T_\theta$ first becomes dependent upon $w$ at the given carrier content $p$. This $T'(p)$ {or $T_{KTB}(p)$} marks the temperature above which the set frequency of operation, $w$, exceeds the correlation dissolution rate, ${}^{1}/t(T,p)$ {$\equiv \Omega(T,p)$}. With higher temperatures it becomes necessary to move up to ever higher probing frequencies if one is to retain the semblance of coherence. For BSCCO-2212 Corson *et al* indeed did uncover such behaviour, determining $T_{KTB}(p)$ to be 60 K in a sample with $T_c$ = 75 K (*i.e.* $p \approx 0.11$ [90]), and $T_{KTB}(p)$ = 15 K in their $T_c$ = 33 K sample ($p \approx 0.07$). In the above description one might note that the standard KTB treatment has been surreptitiously extended from a system homogeneous circumstance to the present inhomogeneous stripe phase condition with its negative-*U* centres. Corson *et al*'s results would appear to support such extension. They give for example the appropriate phase angle evolution in the response from 0 to ${}^{\pi}/_{2}$, of a dissipative to fully lossless status, as the applied frequency $w$ is raised to well above $\Omega'$. For the above $T_c$ = 75 K sample the relevant $\Omega'(T)$ values advance rapidly from $10^9$ to $10^{13}$ Hz as $T$ increases from 60 to 95 K. The upper fluctuation rate here has become identical with the ballistic dynamics rate for the normal state quasiparticles. Thus this 95 K limit demarcates the 20 K span above $T_c$ within which fluctuations may be ascribed to superconductive pairing.

Of course within HTSC materials there persists above such a temperature the chronic intersubsystem scattering which so patently disrupts the proper establishment of Fermi liquid properties [1f]. It is this impairment of the normal state conductivity by $1^1/_2$ magnitudes that would appear responsible now for displacing the under- and optimally-doped HTSC superconductors away from the remarkable 8-decade-wide log-log correlation that Dordevic *et al* [91] have revealed to exist between $l_c$ and $s_{1,n}$ over the widest range of superconductors. These authors regrettably do not examine too closely what the reasons are for the abnormal values of $n_s$ and $s_{1,n}$ in HTSC materials. $n_s$ as noted earlier is much diminished in underdoped material and only with light overdoping do $n_s$ and $s_{1,n}$ start to run up towards standard unperturbed values. Dordevic *et al* duly show the data points for such samples to veer around sharply towards their universal correlation line.

We shall proceed next to examine from the present viewpoint the highly revealing extension which Corson and coworkers have recently made [92] to their above time domain spectroscopy. With the F.S. scattering geometry for the HTSC cuprates configured as is indicated in figure 3 of [1f], there occurs strong quasiparticle scattering out of the 45° directions (nodal as regards the subsequent $d_{x2-y2}$ superconductive gapping) and into the axial (antinodal) {π,0} van Hove saddle regions. Quasiparticles from these high DOS saddles in turn undergo chronic scattering through association into pairs, the partners here being derived from orthogonal M points in our double-loading, *p*/*d* shell-filling fluctuations into the negative-*U* centres. In the latter the requisite (antibonding) excitation energy per pair of quasiparticles is, as we shall find, almost exactly countered by the negative-*U* energy accruing to that pair (see fig. 1 of [1f]). This takes the bosonic pairs back into close resonance with $E_F$ – in fact slightly below once within the condensed state. Now Corson *et al* [92] had been looking to identify the characteristic quasiparticle lifetime within the superconducting state, since there seemed to exist a problem in reconciling the ARPES with the microwave data. Their time domain pulse spectroscopy was judged well suited to resolve this, and in fact did so. It disclosed that the signal crossover being sought (with regard to whether $s_1(w)$ either rises or falls after cooling below $T_c$) is encountered for the case of slightly underdoped BSCCO at an applied frequency of about 0.5 THz. It additionally was shown that $1/t_{qp}$ below $T_c$ follows the same linear-in-*T* relation so characteristic of the normal state and that forms part of the Marginal Fermi Liquid behaviour of HTSC cuprates formulated originally by Varma and coworkers [93]. While resolving one problem Corson *et al* however raised another, because below $T_c$ it proved totally impossible to model the observed sample conductivity within the customary two-fluid treatment. The observed conductivity at 'low' frequencies is much too elevated, just as the microwave work had indicated. By employing the observed spectral weights directly, rather than deduced values for the carrier density fractions ($r_s$ and $r_n$, with $r_s + r_n = 1$), Corson *et al* were able to reveal that for frequencies of less than 1 THz a third very significant contribution to $s_1(w)$ is additionally in play below $T_c$. This extra contribution is not simply due to thermal fluctuation of the superconducting order parameter; such effects are detected separately as a marked peak around $T_c$. The novel contribution evident upon cooling below $T_c$ is found in fact to mount exactly in step with the superfluid density $r_s$, the temperature independent ratio between the two being here 0.3. Corson *et al* suggest that one might look to sample granularity and problems with intergranular Josephson coupling in establishing some collective mode related to the screened plasma frequency of the condensate. While I endorse a system inhomogeneous origin, it seems more in keeping with the present situation that one is uncovering here is a sizeable population of bosons lying outside the condensate and thermally dispersed over the evaluated mode energy width of ¼ THz (≈ 1 meV or 10K). Recall that in superfluid liquid $^4$He more than 80% of those composite bosons reside at any one time outside the condensate [94]. The level of inhomogeneity in the HTSC cuprates is actually so marked it is rather remarkable that as many condensed pairs arise as are found. Even below $T_c$ the quasiparticles reveal by their odd lifetime behaviour how extreme the scattering and trapping conditions are in these materials particularly if underdoped [1d, App A].

We shall return shortly to address the consequences of the above decompostion in relation to the neutron and ARPES results, currently so much argued over. First though it is necessary to back up further our claims to negative-*U* behaviour – and with the large magnitude specified.

## 3. Direct probing of the negative-*U* scenario; progress with optical and phonon measurements, including the phonon scattering of neutrons.

In reference [1e] the work of Holcomb *et al* [95] and Stevens *et al* [96] was developed as prime experimental evidence in support of the current exposition. The former involves very precise thermal difference optical reflectivity (TDR), while the latter uses the laser pump/probe optical technique applied both in reflection and transmission modes. Each set of experiments focusses upon the sub-charge-transfer edge, where following passage below $T_c$ small but interestingly structured spectral changes are known to arise near 1.5 eV. Stevens *et al* discovered employing this energy of probe two distinct populations of pump-excited relaxation products existing with picosecond and longer than nanosecond timescales respectively. The behaviours of these two differ significantly in form, but both 'hot' populations clearly express a relation thermally to the superconducting order parameter. This outcome was presented in [1e] as a further manifestation of the inhomogeneous two-subsystem nature of the HTSC materials, and more specifically of metastable pairs becoming established at (some) negative-*U* centres. Holcomb *et al* forced through a homogeneous Eliashberg type analysis in an attempt to demonstrate that their data would support an electron-boson coupling function based upon the given high frequency absorption band. From their long and careful analysis the coupling constant *l* emerged in the elevated range of 1.35 to 1.50. Each group since has extended its work to additional HTSC materials and has analysed the results more closely [97,98]. Before turning to examine this later work however it is best to introduce first some related results obtained very recently by Li *et al* [99] which are more robust and which carry novel ancilliary information allowing them to provide powerful justification of the present approach.

The technique selected by Li *et al* [99] moves from a purely optical one to one using laser activation of a thin-film electrical waveguide. The latter was 30 μm x 5 mm with a film thickness of just 100 nm. The $Ti^{3+}$:sapphire laser supplied 100 fs pulses of magnitude 5 μJ at a repetition rate of 20 kHz. This technique uniquely accesses the condensed pair population and more specifically yet the laser-induced pair breaking, obtaining an experimental signal which now is at the parts-in-10 level, rather than in $10^4$. The fast optical response of the bridge circuit in fact monitors the time derivative of the kinetic inductance and is proportional to $n_s^2.(\Delta n_s/\Delta t)$ . For a patterned $YBCO_7$ film what was revealed upon scanning the tunable laser between 1.45 and 1.65 eV was a remarkably narrow peak centred at 1.54 eV. This can only be excitonic in nature. What is more it is associated with sharp satellite features spaced from the central peak (B) by 41 meV to the low energy side (C) and by 70 meV to the high energy side (A). The three peaks each hold to their positions as the temperature is adjusted below $T_c$, but their relative amplitudes become greatly modified. Measured relative to B, peak C grows very significantly upon cooling, while conversely peak A dies away entirely. Li *et al* treat the above as in some way referring to standard 'p-to-d charge transfer excitation' incurred within insulating segments of the specimen, this in order to try and accommodate the fact that the width of the observed spectral features is so small. However I would like to read the negatively signed experimental signal, with its 100 ps recovery time, as relating to the disruption of those negative-*U* bosons (${}^{10}Cu_{III}{}^{2-}$) being detached from the condensate. In the disruption process one electron becomes transferred directly into the Fermi sea, whilst the other electron absorbs the excitation of 1.5 eV to acquire ${}^{9}Cu_{III}{}^{1-}$ status [see 1e fig.1]. At raised temperatures the above negative-*U* pair loss process can become coupled into the lattice excitations and so demand the simultaneous emission of a phonon, this accounting for peak A. By contrast at reduced temperatures pair elimination increasingly is able to occur at a somewhat reduced laser photon energy because of a spin-flip energy input of 41 meV shed from the coupled condensate. 41 meV represents the spin pair singlet-triplet separation for the majority Cooper pairs, and accordingly it approximately defines the condensate binding energy in relation to $E_F$. Intersubsystem spin coupling of this form will account for peak C and for its rapid growth in intensity towards low $T$. As befits many-body processes the intensities of all three above features are observed to behave nonlinearly with applied laser power.

The 70 meV or 570 $cm^{-1}$ phonons singled out in the above paragraph are rather special ones for $YBCO_7$, being identified in [100,101] as the top IR-active modes. These are the ones where the basal Cu and O sublattices of the primary coordination unit breathe, modifying Cu-O bond lengths; *i.e.* the modes naturally involved with $d^9$ to $d^{10}$ or $d^8$ count basal charge transfers, as met with in the negative-$U$ pair loss process mentioned above. The c-axis J-T mode is known to soften sharply from 571 $cm^{-1}$ to 569 $cm^{-1}$ upon cooling through $T_c$ and to gain in oscillator strength by more than 50% [100]. In [101] general observations were made about the ready detectability even of basal IR-active phonons in the HTSC 'metals', the authors commenting there on the remarkably low level of screening apparent for these systems. This poor screening suggests that at least for the phonons involved the dynamical charge inhomogeneities coupling to the former develop on an appreciably longer timescale than is set by the phonon freqencies, as was intimated above. The modes which show up particularly strongly in the IR work are those known to have eigenvectors involving $c$-axis atomic displacements – in $(La/Sr)_2CuO_4$ and $La_2CuO_{4+\delta}$ specifically those of the apical O(2) atom and coupled La [102]. The La atom is small and highly charged and sits here loosely in a large 9-fold coordinate site. However the La since it is associated principally with the unoccupied CB is not as important to electron-phonon coupling as are the apical oxygen atoms, $c$-axis movement of which governs the Jahn-Teller splitting of the vital $pd\sigma^*$ Cu-O bands ($d_{z2}$ re $d_{x2-y2}$) around $E_F$. Of course optical work on phonons provides a view only of the zone centre activity, and neutron scattering experiments become necessary in order to elucidate what is happening throughout the body of the zone. This forms a difficult task given the small crystals available and the large number of lattice modes involved – 39 in the case of $YBCO_7$. Furthermore very sizeable TO/LO splittings persist in these poorly screened residually ionic systems and a polarization study is called for. Partial sets of neutron determined phonon dispersion curves are available for LSCO and $YBCO_7$, and comparison of these has been made with those from carrier-free LCO and $YBCO_6$ respectively [103]. At several points there is evidence of appreciable mode modification. A certain degree of anharmonicity arises from the strong Cu-O hybridization and the residual non-local Madelung potentials. Some while ago partially self-consistent LAPW-LDA frozen phonon calculations were performed by Krakauer, Pickett and Cohen [104] and a comparison made with the neutron data. Particular attention was paid to the apical J-T c-axis displacement, c-axis propagating modes. These indeed are able to contribute most effectively towards driving the zone-averaged electron-phonon coupling constant for LSCO up above 1, as was determined from a full Allen and Dynes type treatment. Such coupling might therefore just possibly be taken to account in LSCO for a $T_c^{max}$ of 38 K. It is the subsequent HTSC materials which emphatically demand that a more intricate and novel account of events be sought.

Little, Holcomb *et al* in their more recent TDR spectroscopy work [97] on Tl-2212, extended now down to 0.1 eV, affirm that an upper energy phonon is indeed much involved within the HTSC superconductive coupling process. Upon applying their simplified homogeneous Eliashberg treatment to the new data and retaining a rather standard Coulomb 'pseudopotential' $m^*$ value of 0.15, they now emerge with a two-component coupling constant, $l$, comprised of $l_{ph}$ = 1.01 for the phononic part plus just $l_{ex}$ = 0.36 for the 1.6 eV contribution. They find furthermore that the high $T_c$ value arises only under the combination of terms, not either alone. This concurs with our negative-$U$ understanding of events as being a σ/σ* effect within the Cu-O framework. The completion of the antibonding band in our negative-$U$ fluctuation sees the local coordination unit augment in volume (reflecting the changes observed from $d^8$ $La_2NiO_4$ to $d^9$ $La_2CuO_4$ to $d^{10}$ $La_2ZnO_4$ [59a fig.10]), in addition to a shape change ensuing from elimination of the Jahn-Teller distortion at the electron sourcing $^9Cu_{III}^0$ sites. This coupled structural activity accounts for the development of an isotope effect for $T_c$ [105], notably with underdoped material.

With Jahn-Teller involvement featuring strongly too in stripe phase formation [4a], an added isotope effect for $T^*$ is in evidence above $T_c$ [79,106]. EXAFS results supply a direct record of the structural activity associated with the stripes [107]. For LSCO from inelastic neutron scattering there is known to be a marked $2a_o$ phonon softening from 85 to 70 meV (for $x$ ~ 0.1), which recently has been re-examined by McQueeney *et al* and by Pintschovius and Braden [108]. While such behaviour might just possibly relate to short-range dynamic change involving the LTT structural transformation, it more likely reflects the alternating cell nature developed along a stripe in the charge build up occurring there - what in [4a] was termed 'single density' stripe loading. The above action has to be distinguished from that now reported by Mook and Dogan [109] at lower energy in $B_{1u}/B_{2u}$ basal phonon dispersion and line-width studies, again using inelastic neutron scattering. In this latter work the recorded strong phonon softening refers to the perpendicular orientation, and taking a $q$-vector near $(^1/_4,0)^{2\pi}/_a$ it involves the charge partner to the better known incommensurate dynamic spin 'spotting' sited axially displaced from $(^1/_2,^1/_2)^{2\pi}/_a$ [110,4,2]. Of these two dynamic micro-orderings it is the charge order which appropriately takes precedence, it persisting up to higher temperatures, to higher hole doping content, and to greater system covalence [111] – remember even BSCCO displays charge stripes [80], while, by contrast, magnetic effects already are much diminished in YBCO [112] in comparison with LSCO [113]. The above structural activity, sourced as it is from several directions, is not easy to treat with certainty, and when segments are taken in isolation it offers neither a safe nor a sufficient basis upon which to proceed with any formulation of HTSC (as for example was tried recently by Mihailovic and Kabanov [114] in connection with a J-T polaron scenario). One here is not denying a significant role for electron-phonon coupling within HTSC: a new Fano analysis of Raman scattering phonon line shapes has nicely demonstrated how such coupling tracks $T_c$ in LSCO [115]. It is that as regards HTSC the critical electron-lattice coupling takes on a very specific form, it occurs as Mihailovic and Kabanov envisage away from the zone centre, and, at root, it issues from electronic intersubsystem charge transfer both of single quasiparticles and pairs.

What added support for the present negative-U interpretation of events can be extracted at this point from the more recent pump/probe experimentation and analysis of Demsar, Kabanov, Mihailovic and coworkers [98], work following upon the ground breaking paper by Stevens *et al* [96], and to be viewed in conjunction with my own interpretation of those results in [1e]? Briefly let us revisit the circumstances of these particular experiments. A frequency-doubled $Ti^{3+}$:corundum laser is employed to pump quasiparticles up by 3.0 eV from the V.B. to energies high above $E_F$. These hot quasiparticles are produced in 100 fs pulses and then are permitted to relax, successive pulses being delivered widely spaced at 5 μs. In the relaxation chain there are detected to be produced two well-defined relatively long-lived components, these displaying picosecond and up to microsecond timescales respectively. The nature of these two pump-induced hot populations is next examined and monitored as a function of time by now employing the laser as a much lower intensity secondary probe, re-exciting electrons from the above quasi-terminal conditions but this time at its natural frequency ($\equiv$ 12000 $cm^{-1}$ or 1.5 eV), and recording the changes in optical transmission (at 1.5 eV) of the thin film samples being used (~ 100 nm thickness) that have arisen in consequence of the residual presence of the pumped populations. It is found that the initial e-e relaxation avalanche occurs well within 100 fs of pulse delivery, and ~30-40 secondary electrons are generated per initial pump-excited electron. What is being focussed on with the subsequent 1.5 eV probing are the blockaded populations of (i) hot single electrons and (ii) localized states/pairs, these being situated respectively just above and just within the DOS gap. Such hot species have been prevented under the prevailing energy gapped conditions from rejoining the lower states by the virtual absence of e-e and e-ph low energy relaxation channels. Since these blockaded populations comprise only 1% or less here of the undisturbed carrier content and since the pumping raises the local temperature only by a degree or so, the experiments still fall within the weak perturbation regime for which the lattice and electronic effective temperatures are virtually identical. (Note the energy per 10 fs pump pulse is only 0.2 nJ, which when delivered to a 100 μm diameter spot at a repetition rate of 200 kHz amounts to a low energy flux of just 2 $\mu W/cm^2$.)

Of references [98], [98a] concentrates upon the relaxational behaviour of the picosecond component above, tracking the probe induced optical changes first as a function of sample temperature and then in due course of the degree of underdoping in the $YBCO_{7-\delta}$ system. Conversely [98b] examines overdoped (Y/Ca)$BCO_{7-\delta}$ and discovers that the fast component is comprised there in fact of two qualitatively distinct populations, the slightly longer lived of which continues to show above $T_c$ to some appreciably higher $T^*$. [98c] concentrates on the very long-lived component in the pump-induced population, which though much less studied is surely the one of greater interest. Despite manifesting a metastable localized nature it proves very comparable in population to the faster decaying hot quasi-particle component. [98d] extends the same experiments to Hg-1223 where similar results are forthcoming, although with some revealing variations from the other HTSC materials for which pump-probe data now are available (see [116] for Bi-2212, [117] for Tl-2201 and [118] for Tl-2223).

What is so striking above, as with the original results from Stevens *et al* [96], is that they pin-point the same key resonant energy of 1.5 eV, identified originally by Holcomb *et al* [95] and now highlighted again by Li *et al* [99]. Two quite distinct populations are revealed as being simultaneously (and, as far as selection of laser goes, accidently) monitored at this key energy. The excitations occurring under the 1.5 eV probe have in [1e] been argued to be

(i) the intersubsystem excitation of hot quasiparticles

$$\text{e (hot)} + {}^{8}\text{Cu}_{\text{III}}^{0} \longrightarrow {}^{9}\text{Cu}_{\text{III}}^{1-} ,$$

and (ii) the localized pair disruption

$${}^{10}\text{Cu}_{\text{III}}^{2-} \longrightarrow \text{e (at F.S.)} + {}^{9}\text{Cu}_{\text{III}}^{1-} .$$

Note each process involves the production of the J-T active state ${}^{9}\text{Cu}_{\text{III}}^{1-}$. If this ascription indeed is correct then it directly affirms the magnitude of $U_{\text{eff}}$ as being -1.5 eV per electron or -3.0 eV per pair (see fig.1 of [1e]) – precisely as is required to take the pair back into resonance with $E_{\text{F}}$.

## 4. Theoretical matters relating to the ARPES and neutron scattering data, to spin fluctuations and to electronic homogeneity.

### 4.1 Norman *et al*'s recent analysis of the peak/dip/hump ARPES structure.

It is widely acknowledged that the Fermi liquid behaviour of the HTSC materials is at best ′marginal' in the normal phase, and its improvement within the superconducting phase is less than total even for highly overdoped samples. In the normal phase ARPES finds the Fermi cut-off for optimally doped material to be rather poorly defined, particularly in the region of the vHs saddles. Below $T_c$ the photoemission spectrum from the saddle region is characterized by a downward shifted edge with following strong peak, dip and broad hump, these falling respectively at about $1\Delta_{sc}$, $2\Delta_{sc}$, $3\Delta_{sc}$, and $4\Delta_{sc}$ (taking $\Delta_{sc}$ in BSCCO-2212 as ≈ 25 meV). Some say that the main peak diminishes towards and vanishes at $T_c$ following a BCS-like order parameter, others that it broadens out rapidly and dissolves away somewhat above $T_c$. To resolve this point is important because it throws attention either simply upon the spectral weight associated with the peak's appearance or conversely upon the severe scattering by which HTSC generation is propelled. All our previous perceptions of what is going on in HTSC support the latter approach. Since both normal and superconducting states see severe scattering and a significant loss of quantum coherence, it becomes hard to make any clean separation/ascription of the various spectral features present, particularly in the $(0,\pi)$ ARPES spectra. Recently in view of this Norman and coworkers [119] very sensibly have attempted to allow each anomalous EDC spectrum taken as a unit to speak for itself. The experimental photocurrent spectrum enables $\mathrm{Im}\,G(\mathbf{k}',w)$ below $E_{\text{F}}$ to be extracted, and then a Kramers-Kronig treatment of this over the affected spectral energy range (up to 1/3 eV) enables the corresponding $\mathrm{Re}\,G(\mathbf{k}',w)$ to be obtained. From $G(\mathbf{k}',w)$ the associated self-energy $\Sigma(\mathbf{k}',w)$ components to the many body effects can then be calculated. These are relatable back through the spectral function $A(\mathbf{k}',w)$, where

$$A(\mathbf{k},w) = \frac{1}{\pi}\cdot\frac{\mathrm{Im}\,\Sigma(\mathbf{k},w)}{(w - e - \mathrm{Re}\,\Sigma(\mathbf{k},w))^2 + (\mathrm{Im}\,\Sigma(\mathbf{k},w))^2}\ ,$$

to the photocurrent, $I(\mathbf{k},w)$, when appropriate conditions are imposed on the matrix elements, the particle-hole symmetry and the background signal. In their figure 2 Norman *et al* demonstrate that in the vicinity of $T_c$ the signal is a combination of (i) a strong intrinsic incoherent background, (ii) a 'Fermi liquid' component, and (iii) a persistent quasiparticle pairing term. This is in line with the type of admixture encountered in the preceeding sections within our two-subsystem negative-*U* approach. The Fermi liquid component is evident as an $w^2$ contribution to $\mathrm{Im}\,\Sigma(w)$, and an $w$ form to $w - \mathrm{Re}\,\Sigma(w)$ at very low energy. $\mathrm{Im}\,\Sigma(w,T')$ shows a peak at zero energy, a dip near the peak in the corresponding EDC, and then a hump near the dip within the latter. Note $w - \mathrm{Re}\,\Sigma(w)$ reduces to zero at $w = 0$ and peaks sharply at an energy between zero and $1\Delta_{sc}$. Increase of temperature quickly smoothes out all the above features, so that by 150 K $\mathrm{Im}\,\Sigma(w)$ simply shows a monotonic fall away to high binding energies. A similar thermal in-fill of the features occurs for the $w - \mathrm{Re}\,\Sigma(w)$ plots. The full *T* set of $\mathrm{Im}\,\Sigma(w,T)$ curves displays a common crossover point at the spectral dip energy (~ 72 meV), while the $w{-}\mathrm{Re}\,\Sigma(w,T)$ set possesses such a point at the spectral peak energy (~ 48 meV).

In order to analyze the detailed thermal behaviour of the various contributions above, the self-energy was expressed in [119] in quasiparticle (coherent and incoherent) plus broadened BCS-like pair form by -

$$w - \mathrm{Re}\Sigma = w\left[1 - \frac{\Delta_r^2}{w^2 + \Gamma_r^2}\right]$$

and

$$-\mathrm{Im}\Sigma = c + c_{F.L.}.w^2 + \left[\frac{\Delta_i^2\Gamma_i}{w^2 + \Gamma_i^2}\right]\ ,$$

where $\Delta_r$ is the above crossover energy for the real component, associated with the peak in *A*, and $\Delta_i$ " " imaginary " " dip in *A*.

Note that $\Delta_i/\sqrt{Z}$ is approximately the same as $\Delta_r$, *Z* being the mass factor and reciprocal of the quasiparticle renormalization factor, *z*. The chronic electronic scattering rate involved with pair production and loss is monitored by the $\Gamma$'s.

The results which Norman *et al* pull out of this analysis are most appropriate. *Z* climbs to a peak value of 2 at $T_c$ and then quite quickly settles back to somewhat below 1 by 140 K. $c_{F.L.}$ starts from a *T*-independent value well below $T_c$, drops strongly through $T_c$ and already is tailing away towards zero by 140 K. The incoherent background term *c,* from a raised 60 meV temperature independent base, increases sharply through $T_c$ and then follows a slower linear rise above. The pair self-energy widths $\Delta_r$ and $\Delta_i/\sqrt{Z}$, from a common temperature independent value below $T_c$, decrease steadily above $T_c$. $z$ (i.e. $1/Z$) displays the inverse behaviour diverging from a constant value of 0.4 below $T_c$ to reach a value of 1 by 105 K. Through all the above the spectral peak energy, sited at the energy where $w - \mathrm{Re}\,\Sigma(w)$ passes below zero, emerges as virtually unchanged at ~ 50 meV until the feature finally becomes indiscernable around 125 K. The impression often voiced is that the peak energy drops to zero at $T_c$ (90 K in the present near optimally doped sample). By contrast what is seen here is an ever accelerating broadening of the peak spectral weight: together rates $\Gamma_i$ and $\Gamma_r$ increase with $T$ in power law fashion right from $T$ = 0 and showing no discontinuity at $T_c$ (where $\Gamma$ ~ 7 meV). Various further measures of the spectral peak width indicate a sharply divergent behaviour above $T_c$. Some growth in spectral weight in the residual peak possibly arises above $T_c$, and this Norman *et al* suggest might be understood as the superconducting pair gap above $T_c$ becoming superceded by the RVB spin pseudogap.

All these results appear in line with our two-subsystem negative-$U$ approach to HTSC, and above $T_c$ with some marginal Fermi liquid condition left in the system. Incoherence develops swiftly above $T_c$, and with it a resistivity that ultimately breaks away from $k$-space centred Fermi liquid form into diffusive directly thermally controlled dynamics, as $r(T)$ is carried upto and beyond the Mott-Ioffe-Regel limit. This disintegration of the Fermi liquid has been demonstrated in previous ARPES studies as being most advanced at the vHs saddles and from there to extend out progressively to the rest of the F.S. [42]. In figure 3 of [1f] it was indicated how even the lighter carriers of the [110] directions are subject to strong scattering into the saddle-point sinks. At lower temperatures pairs of carriers from neighbouring sinks are increasingly drawn off into the negative-$U$ shell closure transfers in the minority subsystem. Only however with the close approach to $T_c$ does an appreciable fraction of the pairs become organized into a coherent condensate, as the full Bose behaviour is seeded and propagated by pair injection through into the majority subsystem. Such dynamic exchange causes the above $\Gamma$'s to diminish towards zero only algebraically (approximately quadratically) as $T$ tends to 0 K.

### 4.2 Dispersion kinks in ARPES results and collective bosonic modes.

We have seen from Corson *et al*'s recent work [92] that there is clear support for a substantial bosonic pair population below $T_c$ which actually resides outside the superconducting Bose condensate. Moreover this population mounts on cooling in step with the growth in the superconducting order parameter. These uncondensed bosons will be thermally dispersed in energy and spread in momentum by the full range of collisions. Being of energies close to the chemical potential and suffering boson/fermion interconversion and strong Coulombic interaction, the bosonic collective mode which emerges must interact strongly with the quasiparticle band wherever the two **k**-space structures meet. This will come at roughly the maximal superconducting band gap value below $E_F$. Since the interaction is direct and electronic, the self-energy $\Sigma(\mathbf{k},w)$ of the quasiparticles will be substantially modified there. Recently, in high resolution ARPES work on Bi-2212, a marked kink ~ 50meV below $E_F$ has in fact been recorded in the quasiparticle band dispersion for the [kk0] nodal direction by Bogdanov *et al* [50], Kaminski *et al* [45] and Valla *et al* [48,49]. In this superconductively ungapped direction the band velocity at $E_F$ becomes appreciably reduced by the kinking, with the associated upward band mass renormalization factor being ~ 1.5, much as might be expected from electron-phonon interaction.

Classical superconductors function of course through el-ph mediation, and recent ARPES work [120] on the low temperature superconductor Mo has directly affirmed this, detecting a strong change in $\Sigma_1$ and $\Sigma_2$ at Z.B. phonon energies. Repeating the experiment upon the layered superconductor 2H-$TaSe_2$ reveals however something rather different [121]. Again a kink has been recorded, this time in the EDC dispersion for the inner $\Gamma$-centred piece of F.S., signalling thereby a strong, self-energy modifying, modal interaction at 65 meV. Here now though this is almost double the top phonon energy. What is more (and as for the HTSC case) this kink/self-energy feature, which first emerges below 100 K, *strengthens* upon further cooling, not vice versa as would be the case from a phononic sourcing. The kink does not in fact arise in this case from phonons but from the phasons and amplitudons that relate to 2H-$TaSe_2$ bearing a CDW below 110 K [122]. The part of the Fermi surface which actually is being reported on in this 2H ARPES work is not that directly involved in the CDW generation. The latter is driven by the nesting of a second closed segment of the F.S. centred about each zone corner [123]. The above bosonic excitations of a CDW first were formulated by McMillan [124]. These, being dispersed throughout the entire zone, proceed to affect the residual 'passive' part of the F.S. in 2H-$TaSe_2$ in the manner now noted. Below $T_o^{CDW}$ $\Sigma_1(\mathbf{k'},w)$ becomes increasingly strongly peaked at an energy climbing to 65 meV, while below this peak $\Sigma_2(\mathbf{k'},w)$ steadily descends from its higher binding energy plateau. These changes become reflected in a dramatic improvement in lifetime for the 'inner surface' fermions, as carrier scattering from the ARPES-studied (lighter) segment into the outer saddle-point sinks that govern the CDW formation becomes prevented by the growing F.S. gapping there.

This behaviour in 2H-$TaSe_2$ is so comparable to what now is being observed in the HTSC materials that it fully endorses the suggestion made earlier by Norman, Ding and coworkers [41] that the ARPES peak/dip/hump structure, traceable far from the F.S. and dispersing weakly back towards $\Gamma$, proceeds from the action of a bosonic collective mode. This mode we have already asserted is not of phonons nor indeed magnons - it has the wrong thermal behaviour, but is formed from the great many negative-*U*-sourced charged bosons existing outside the $\mathbf{k} = 0$ Bose condensate. Whether the striped geometry configuring the two-subsystem whole might play some appreciable role in structuring this mode and its scattering fluctuations is a matter for further detailed study. In the HTSC materials, just as in 2H-$TaSe_2$, the present effects do not terminate entirely once above the onset of the low temperature LRO, but project in diffuse fashion to somewhat higher temperatures, particularly where encouraged by suitable defects and impurities.

Now it is actually observed for the HTSC materials that the major kink to be developed in the nodal dispersion curve commences considerably above $T_c$, and indeed the key augmentation to the kinking which sets in at $T_c$ at first went undetected [e.g. 47, 48]. It required the extensive use of synchrotron radiation to bring out this modification in coupling of the collective boson mode to the quasiparticle dispersion curve once below $T_c$. Although the limit to the overall dispersion change occurs at $w_c \approx -230$ meV, the additional effects associated specifically with the establishment of superconductivity extend only to –100 meV, in optimally doped Bi-2212 the kink energy itself being $\approx -60$ meV. The overall kinking, or equivalently the peaking in $\mathrm{Re}\,\Sigma(w)$, displays a magnitude which diminishes steadily as the level of sample doping (i.e. of delocalization and screening) is advanced. This is reflected by a *monotonic* reduction with $p$ in coupling strength between the q.p. holes and the collective mode assessed in [49] by the coupling parameter $l = (\partial \mathrm{Re}\,\Sigma(w)/\partial w)_{E_F}$, the small momentum dependence of $\mathrm{Re}\,\Sigma$ here being neglected. $l$ is found to decrease from 1.7 for an underdoped sample with a $T_c$ of 65 K to only 0.4 in an overdoped sample of the same $T_c$. This steady fall-off in the underlying coupling is to be contrasted with the behaviour witnessed in the additional kinking more immediately associated with the mode and superconductance itself. These extra effects from the bosonic mode are monitored in [49] by taking the difference between each plot of $\mathrm{Re}\,\Sigma(w)$ for $T < T_c$ and one for some $T$ well above $T_c$. Two parameters actually are extracted in [49] to describe the low energy regime as a function of doping; namely $w_o^{sc}(p)$, which designates the binding energy at maximal difference between each above pair of $\mathrm{Re}\,\Sigma(w)$ curves, and, less focussed, $w_o^{max}(p)$, for which the net maximum value of $\mathrm{Re}\,\Sigma(w)$ occurs. Since $p$ for '$Bi_2Sr_2CaCu_2O_{8+\delta}$' is not readily extractable (it depending upon the true Bi/Sr balance in addition to $\delta$) these parameters were finally displayed as a function of $|T-T_c|$. It is encouraging to find that the energies $|\omega_o|$ climb to a maximum from either side of optimal doping and that they do not display the monotonic decline from the underdoped condition shown by $l(p)$ above. The linear relation found between the $w_o(p)$ and $kT_c(p)$ (of gradients 7.4 and 5.1 for $w_o^{sc}(p)$ and $w_o^{max}(p)$ respectively) expresses the same form of relationship to the order parameter that is exhibited by the neutron resonance peak [32,37,39]. Indeed Valla *et al* [49] discover a virtually identical thermal scaling for these two very different physical entities. In underdoped samples the magnitudes of both the nodal $\mathrm{Re}\,\Sigma(w)$ peak and the $(\pi,\pi)$ neutron peak fall most steeply at $T_c$ and they finally vanish by ~ $T_c$+25 K: each after their own fashion provides a record of the thermal development of the superconducting order parameter. Valla *et al* [49] relate all this to spin fluctuations in some magnon mode, because, with a kink onset energy $w_c \approx 230$ meV, the full energy range affected in the ARPES data is of the order of $2J$ in more metallized samples. This line has been adopted too by Eschrig and Norman [125], largely in the absence of any contemplation by them of the negative-$U$ option. However, as Valla and coworkers point out, the disparity in behaviour with underdoping displayed by $l$ and $w_o^{sc}$ demonstrates that while this mode coupling (as with magnetic effects) is strongest in underdoped material, the maximization of $T_c$ and more particularly the condensation energy is dependent upon physical circumstances which require a higher degree of delocalization, as witnessed in the specific heat results of [23]. Furthermore whatever occurs in the $(\pi,\pi)$ direction near $E_F$, or indeed at $(\pi,\pi)$ itself, this is at some remove from the actual expression of the superconductive gapping principally incurred in the vicinity of (π,0). *l* indeed is observed to rise as one moves away from the nodal direction.

As one migrates from the node towards the saddle what actually is found is that the kink develops into a clean break in the dispersion curve at around 80 meV. Steadily the increasingly detached lower energy segment of the associated EDC becomes detectable across more and more of the zone, settling into the strong, narrow, virtually dispersionless form previously discussed in regard to the peak, dip and hump structure of the M-point EDC spectrum. This strong peak occurs very close to 40 meV; i.e. at somewhat less than $2\Delta(0)^{max}$. The full spectral progression has been laid out in detail in figure 4 of [45], where the above behaviour for the superconducting state is to be contrasted with that found up at 140 K. It is this data set that Eschrig and Norman in [125] set out to model. Functionally they accomplish this by taking the bosonic activity to exhibit marked enhancement for **q** near **Q** = (π,π) of the form

$$f(\mathbf{q}) \;=\; \frac{c_Q . x^{-2}}{x^{-2} + 4\left(\cos^2(q_x/2) + \cos^2(q_y/2)\right)}$$

where $x \leq 2a$, and taking $c_Q$ from neutron spin scattering [37] to be $0.3\mu_B^2$. The full susceptibility

$$c(\mathbf{q},w) \;=\; -\,f(\mathbf{q}).\left[\frac{1}{w - \Omega \pm \mathrm{i}d} \;-\; \frac{1}{w + \Omega \pm \mathrm{i}d}\right]$$

has Ω, the frequency of the interactive boson mode, taken to be q-independent and set to 40 meV.

Ever since Pines and colleagues' work [126], and more recently as extended by Abanov, Chubukov and coworkers [127], such a form to the generalized susceptiblity has been widely attributed to antiferromagnetic spin fluctuations. It is this ascription that I have claimed needs recasting in terms of negative-*U* shell-closure charge fluctuations. These too are associated preferentially with (π,π) as explained in [1f]. The superconductive quasiparticle gapping becomes dominant at the M points, and the mode coupling is strongest there, this being in consequence of the susceptibility to *on*-coordination unit spin-singlet pair formation, not *between*-coordination unit antiferromagnetic spin singlet structuring. The effective coupling constant *g* empirically settled upon at 0.65 eV in the analysis by Eschrig and Norman should not be viewed as setting the scale of events to magnetic levels. For *x* so small only charge effects are sufficient to uphold the experimentally observed phenomena.

### 4.3 Spin fluctuations and the Marginal Fermi Liquid.

With regard to the spin fluctuation scenario of HTSC it always has seemed rather improbable as to why, if this is all there is to the mechanism, HTSC ought not to be considerably more widespread. Where are the vanadium and titanium HTSC systems? We note too with $Sr_2RuO_4$ how spin paramagnon coupling supports there only a very small $T_c$ ~ 2 K [128]. While for the cuprate systems very considerable fitting achievements have been made by Chubukov and colleagues [127], not only of the ARPES and neutron data but also the d.c. and optical conductivities, one cannot help feeling that these have been gained at a price. With reference specifically to [127c], is the magnetic coupling really so strong with $x$ so short? Can the spin fluctuation frequency really be so low? – viz. $w_{sf}^{(\text{saddle or 'hot spot'})} \approx 5$ meV ($w_{sf}$ is here the crossover frequency between Fermi liquid behaviour ($\Sigma_2 \propto w^2$) and magnetic quantum critical behaviour ($\Sigma_2 \propto w^{1/2}$)). Are the deduced forms for $\Sigma_2(w)$ and $\Sigma_2(T)$ truly as linear as is being asserted? – and correspondingly from these the derived photoemission peak width and d.c. resistivity? Where do the very substantial constant terms added onto the optical $s_1^{-1}(w)$ and photoemission line-shape expressions derive from? Above all why are the HTSC cuprates treated as structurally and electronically homogeneous when their two-subsystem form is so well established? The whole focus on the primacy of spin-fluctuation behaviour, centred around a susceptibility strongly peaked at $\mathbf{Q} = (\pi,\pi)$, currently hinges upon precisely how one interprets the neutron spin-flip resonance peak, a feature that develops only below $T_c$+25 K [30]. This particular peak surely is to do essentially with superconductive pairing and not with antiferromagnetic coupling. Quasi-local singlet spin pairing, its creation and destruction by electronic as distinct from phononic means, is not the sole preserve of magnetic coupling or excitation. Any composite bosonic mode falling close to $E_F$ and suitably dispersed may potentially mimic what is claimed to arise within the spin-fluctuation scenario, particularly when its sourcing is centred upon the same $k$-space 'hot spots' – as presently is the case.

With the HTSC cuprates it is clear that the non-Fermi liquid state exhibited, in particular above $T_c$, bears the mark of a very unusual and specific origin. Even the mixed-valent vanadium compounds $LiV_2O_4$ and $NaV_2O_5$ do not hold to the Marginal Fermi Liquid expressions introduced early on by Abrahams, Varma and colleagues as being relevant to the HTSC cuprates [93]. $LiV_2O_4$ is likened now to a Heavy Fermion compound [129]. Its high electronic specific heat is quite dissimilar to that of YBCO, etc. [23], and has recently been argued to follow from low energy excitations in the quasi-1D $s = {}^1/_2$ subsystem [130]. $NaV_2O_5$ at 34 K displays charge ordering and a weak $s = {}^1/_2$ spin-pair dimerization, from a condition which above $T_o$ is site equivalent towards the NMR probe [131]. No superconductivity shows up here, and that which does so in $Li_{1+x}Ti_2O_4$ is at a quite modest 12 K [132]. $(La/Ba)TiO_3$ likewise is a system of rather poorly defined stoichiometry, Pauli paramagnetic and weakly metallic, but now with no sign of superconductivity, while $LaTiO_3$ itself is slightly distorted, non-metallic and antiferromagnetic [133].

Abrahams and Varma [134] recently have revisited the origins to the prescribed functional form of their MFL. This analytically constitutes a very particular first step in the breakdown of the $T = 0$ Fermi liquid density of states discontinuity. The original formulation of

$$\mathrm{Im}\, c(\mathbf{q},w,T) \begin{cases} = -N_o.(w/T) \;, & w \ll T \;, \\ = -N_o.(\mathrm{sgn}w) \;, & T \ll w \ll w_c \;, \end{cases}$$

translates through into the self-energy expression

$$\Sigma(\mathbf{k},w) = l[w.\ln(x/w_c) - i.\pi/2.x] \;,$$

where $x \approx \max(|w|, T)$ – as for example $x = \sqrt{(w^2 + (pT)^2)}$ since noted in neutron scattering [29]. $l$ is the many-body coupling constant. Unlike for standard metals (i) $\Sigma_2$ has become a very sizeable fraction of $\Sigma_1$, (ii) there is the unusual $w/T$ form to $\mathrm{Im}\, c$ at low frequency, (iii) $\Sigma_2$ varies as $w$ not $w^2$, and (iv) most notably there is virtually no momentum dependence to events. The **k** independent behaviour here is quite unlike that seen at low temperatures once experiencing superconductive coupling and gapping. Since at $T = 0$ and $w = 0$ (i.e. at $E_F$) $\Sigma_1 \propto w.\ln w$, $1/z \equiv (1 - \partial\Sigma_1/\partial w) = -\ln w$, and therefore $z$, the quasiparticle residue, must drop to zero. We have then a form of localization, as has long been implied by the $1/T$ behaviour of the Hall constant [1d]. The superconductivity at low temperatures masks the visibility of this behaviour in the electrical resistivity, but in the thermal conductivity we can see something of this limit because as we are dealing with *d*-wave gapping the nodal electrons still make a significant contribution. Thermal conductance data from YBCO-124 by Hussey *et al* [135] demonstrate that these nodal quasi-particles are rendered non-metallic at low temperatures. Comparable circumstances have been registered in HTSC systems in very high magnetic fields under which the superconducting gap is partially closed [136]. Working in such fields is however a strong perturbation on many factors and the electronic thermal conductance experiment on 124 is more definitive. Weak localization often arises through structural disorder but YBCO-124 is a stoichiometrically well-defined compound, with crystallographically no immediate register of its mixed valence; indeed NMR on 124 would indicate a single planar Cu site signal [137]. That signal however is broadened somewhat and strong quadrupolar charge effects are detected. In the light of our two-subsystem approach to HTSC (N.B.124 has a $T_c$ of 80 K, rising sharply under pressure [138]) this encourages one to probe further. Directly one does so one discovers by means of the PDF neutron technique that in fact the structure at the micro-level is severely disturbed [139]. As anticipated this manifests itself in J-T apical oxygen shifts, together with various knock-on consequences in the chains where oxygen atoms become displaced transversely. Here then is the anticipated dynamic charge activity that so characteristically shapes the normal state and upholds HTSC. The time scale of this charge activity and its structural response is what dictates the steady growth in non-Fermi liquid behaviour. Something rather similar occurs in the R.E. interconfiguration fluctuation materials like $SmB_6$, SmS and TmSe [140]. In the Sm case the configuration fluctuations involved there are between $f^6$ and $f^5d^1$, and the very marked difference in ion size between these two limiting conditions results in very strong coupling to the lattice. Strong localization ($\log r \propto T^{-1/4}$) ensues in the configuration-mixed ambient conditions [141]. There UPS, operating at $10^{-15}$ s, registers simultaneously the characteristic *f*-electron spectra appropriate to *both* terminal configurations, whilst the much slower Mossbauer probe (~ $10^{-6}$ s) detects just a single quantum mechanical combination. Strong pseudogapping in the FIR is observed in such systems [142]. Correspondingly in systems experiencing a frustrated disproportionation or CDW, comparable freezing of the low frequency optical conductance can again be seen, as for example in $(Ba/K)BiO_3$ [143], $K_3C_{60}$ [144], and very recently 9L-$BaRuO_3$, a trimer chain structure [145].

What encouraged Abrahams and Varma [134] to return to examine the cause of MFL behaviour was that the new high resolution ARPES data, and in particular the MDC cuts of the ($E$,**k)** dispersion curves [47-49], now permit a detailed line-shape/line-width analysis. Throughout the zone near $E_F$, $\Sigma_2$ is shown to be proportional to $x$, and scarcely dependent on $\hat{\mathbf{k}}_F$, with no evidence of F.L.-type $T^2$-dependence anywhere, even at the nodes. What was not originally foreseen is that for a given temperature $\Sigma_2(w)$ acquires the form $\Gamma(\hat{\mathbf{k}}_F) + l(\hat{\mathbf{k}}_F)w$ , with the constant term, $\Gamma$, always large, in particular at the saddles. $\lambda(\hat{\mathbf{k}}_F)$ conversely displays a weak maximum at the nodes. Because the $\Gamma$ term is virtually independent of frequency and temperature it has been attributed by Abrahams and Varma to static impurities. However, as we saw above, it is misleading to identify such scattering as 'impurity' scattering. The responsible structural disturbances are intrinsic to the mixed-valent cuprate systems, and they are central to establishing the two-subsystem negative-$U$ behaviour there. One has structural inhomogeneity and the concomitant electronic inhomogeneity in all these mixed-valent oxide systems, whether their non-integral electron count is being obtained by excess oxygen insertion as in BSCCO [146] or by counter-ion substitution as in LSCO [147]. The intrinsic, homologous nature to the critical inhomogeneity is similarly manifest in YBCO-124 as we have seen, even though as averaged over time and space it would appear structurally regular. Charge/site inhomogeneity taken singly, or indeed in the form of stripes, is essentially a dynamic condition, and as such does not tie HTSC to be dependent on any but the most general of structural circumstances. Long ago we observed how very similar $La_2CuO_{4+\delta}$ is to $(La_{2-x}Sr_x)CuO_4$ in the matter of its superconductivity [148].

Not only is this question of local mixed-valent structural inhomogeneity absent from Abrahams and Varma's considerations, but it is the element that critically is absent from all mean-field spin-fluctuation and band structurally based theoretical work [149], and indeed from empirical interpretation of much of the intimately connected neutron spin scattering data. The latter in fact carry strong intimation of the charge domain structuring which so often has been neglected (as in the residual linewidth $1/\kappa_o$ analysis surrounding figure 4 and equation 5 in [29]). One of the few theoretical papers fully to embrace the importance of the structural meso- and micro-scopics is that appearing recently from Chakravarty and Kivelson [150], which relates not only to the cuprates but to doped $C_{60}$ and the polyacenes also. In the latter cases it is acknowledged how 'chemical' pair binding and the influence of particular electron counts (in regard to disproportionation, etc.) are crucial towards countering the highly repulsive character of the positive Hubbard $U$ term inevitably in play in all tight-binding systems.

**4.4 The band structurally based HTSC deliberations of Gyorffy and colleagues.**

This body of research has taken the line that since density functional theory (DFT) in band structural work implicitly incorporates all the many-body interactions in enumerating the ground state energy, the extension of DFT to the Bogoliubov-de Gennes (BdG) formulation of superconductive pairing will yield information on that pairing through an appropriate parametral accommodation to the experimental data [151]. Because of powerful new numerical methods it has been possible to carry through such DFT/BdG analysis on a TB-parametrization of a high quality LDA band structure calculation [152], concentrating on the crucial $CuO_2$ planar states which lie within ± 2 eV of $E_F$. The pair interactions considered are, as with the BCS interaction, very local in nature, and accordingly are in turn open to TB parametrization in a fashion similar to the band structure itself. The classic BCS interaction is of retarded $r = 0$ delta-function form, but now, in view of the small $x$ value, it seems more appropriate to consider non-retarded on-site and nearest-neighbour options. In [151] half a dozen different pairing possibilities were tested out. The manner of doing this was to fix $T_c^{max}$ at the experimental value and then to hunt down that pairing channel for which the BdG kernel is smallest in magnitude to yield this; in this way one identifies the most proficient pairing channel. Almost inevitably from the options considered, intra-layer nearest-neighbour Cu $d_{x2-y2}$ - $d_{x2-y2}$ coupling emerges as being the most effective mode, with a kernel energy parameter, $K_a$, of just 0.68 eV. Unfortunately this option, which goes forward to cover moderately well the photoemission gap, specific heat and penetration depth data [151,153], was not investigated further within the context of the site-inhomogeneous, three-centre, mixed-valent format presented in [1] and the sections above. This presumably might be achieved through the CPA alloy procedure. As was noted in our discussion of the 124 compound, even the most regular of the HTSC systems in fact is structurally and electronically inhomogeneous on a time scale long compared with most electronic processes [138,139]. Hopefully the above extension would lead to an even smaller kernel magnitude being determined. Perhaps as importantly it should lead to a better understanding of the real significance of the vHs in generating the high $T_c$ of the cuprates, in addition to the near-parabolic form which so characterizes $T_c(p)$ [90].

The above series of papers [151-153] begin actually to fall into the trap of arguing that the rise and fall of $T_c$ simply is a DOS/vHs effect, rather than that the kernel itself is doping dependent. Of course this is because there is nothing explicit within the employed LDA band structural approach relating to the decaying away of magnetic spin effects and the advancement in delocalization as one moves significantly from band half-filling (i.e. uniform $d^9$ occupancy). A major effect of the vHs is to make carriers heavy within the vicinity of the M points $\{0,\pi\}$ in the zone. To my understanding this is what for local pairing favours a $d_{x2-y2}$ symmetry order parameter being established under the above pairing kernel. The M points are where the quasi-particles are preferentially drawn from into the shell-filled, negative-$U$ state. It is of value to question what is the most favourable energy of the saddle-point relative to the Fermi level. We attempted to do this in a recent simple BdG modelling [154] of the effects which the precise shape and position of the normal state DOS have upon $\Delta$ and $T_c$. It was felt, and indeed appears the case, that it is preferable for the weight in the DOS to be shifted down somewhat below $E_F$, rather than to be placed right at $E_F$ as so often is presumed. Interestingly the best LDA band structure calculations always site the vHs a little way below $E_F$, and this is endorsed in photoemission work which registers a strong feature at such a binding energy, frequently referred to as the 'extended saddle point' [155]. The latter extension is just conceivably as much to do with the photoemission process as it is with the saddle geometry itself – recall the carriers there are heavy and prone to disorder localization. Calculation suggests that the saddle energy seems to track $E_F$, each falling as the number of band electrons decreases (*i.e.* number of holes rises) [156]. In fact there may be some interesting correlation effect here tieing the two together [157]. Certainly there is to be found no indication from specific heat work that with decrease in electron count the Fermi level ever descends through a DOS maximum [23].

This leads one to query why for underdoped HTSC systems $T_c$ under high pressure almost uniformly first rises and then falls – see [158] for the prime case of YBCO-124. The above would suggest that rather than pressure driving the vHs through $E_F$, one primarily is modifying the size of the interaction kernel {treating these two to first order as separable, as in BCS – $T_c = q_D.\exp(-1/N(0)V)$}. One can postulate within the presently proposed scenario that the initial effect of pressure on an underdoped system is from decrease in the *a* parameter to entrench RVB behaviour more strongly and so diminish pair breaking. Ultimately $T_c$ peaks however because with added pressure the electronic system becomes too delocalized to uphold strong local pairing. One may again see here why such notions are not too welcomed by theorists because the number of free parameters is much less constrained. However if the constraints currently being imposed are artificial then the outcome can be highly misleading.

### 4.5 Terminology and treatment of valence and electronic configuration in a mixed valent system and the case of YBCO$_x$.

Speaking of misleading constraints and terminology, some attention needs to be paid at this point to a recent publication from Temmerman *et al* [159] that has direct bearing on the mode of site configuration and valence designation employed in my own papers. I have adopted standard chemical practice in labelling $KCuO_2$ and $LaCuO_3$ as $Cu_{III}$ (or Cu(III)) compounds. This avoids the extreme ionic form $Cu^{3+}$. The latter suggests a tightly bound, atomic-like $d^8$ residue, while the former mode shuns specifying a precise configuration, which in a solid is of mixed parentage due to covalent hybridization with the surrounding ligand states ($pd\sigma/\sigma^*$; $pd\pi/\pi^*$, etc. in M.O. designation). $Cu_{III}$ is a general state/electron counting device, indicating a upper 'valence band' complement of 8 occupied non-bonding ($\pi/\pi^*$) and antibonding ($\sigma^*$) states per Cu centre present. In this representation $VO_2$ and $VF_4$ are $V_{IV}$ compounds – note it is immaterial here that the former compound lies to the metallic side of the Mott transition and the latter to the insulating [59]. Correspondingly both LaS and $LaI_2$ are $La_{II}$ compounds - in this case both metals; likewise GdS and $GdI_2$ are metallic $Gd_{II}$ compounds [160]. These two pairings, each divalent in our standard terminology, take approximate metal atom configurations of $f^0d^1$ and $f^7d^1$ respectively, and they are to be distinguished from trivalent $La_{III}$ and $Gd_{III}$ in say $LaI_3$ and $GdI_3$, where we have outer configurations approximating to $f^0(d^0)$ and $f^7(d^0)$ respectively. I labour this point because considerable confusion has grown up around the so-called 'mixed-valent' R.E. systems [140] in regard to valence and configuration. The proper terminology for $SmB_6$, SmS, TmSe, etc. is interconfiguration fluctuation (ICF) materials. The mix up has occurred because not infrequently GdS and $GdI_2$ are seen referred to as 'trivalent' compounds, in view of the delocalization of their outer *d*-electron. The same applies to the 'collapsed' or 'gold' form of SmS, effectively $f^5d^1$ in its high pressure limit. However nobody would ever refer to $VO_2$ as being pentavalent! – as with $SnO_2$ and $TiO_2$ it is quadrivalent. The 'residual' d-based state in question is not of primary bonding σ-type geometry. Metallic $Sm_{II}$ $f^5d^1$ SmS and insulating $Sm_{III}$ $f^5d^0$ $Sm_2S_3$ are of different standard valencies. The valency essentially is set by the ligand stoichiometry, although care must always be taken to account for any strong direct X-X bonding causing the ejection of the complementary empty antibonding X-X states up out of the 'valence band'. It emerges in this fashion that $NiS_2$ (with its X-X pairs) is divalent like NiS, and so too is $NiP_2$ (with its Se-like chains) [59]. In the main with HTSC systems the ligand sublattice is straightforward, although electron counting problems from this cause have been observed to arise in certain excess oxygen insertion materials, e.g. $Hg_xBa_2CuO_{4+\delta}$ [20].

Now in a truly mixed-valent compound such as $Sm_3S_4$, specifying the presence of $1Sm_{II}$ and $2Sm_{III}$ immediately will convey the correct electron numerology. In $Sm_3S_4$ we are in fact dealing essentially with the purely ionic *f*-electron quasi-core-like configurations of $f^6$ and $f^5$ respectively. However upon turning one's attention to $Fe_3O_4$, and particularly to $Fe_3S_4$, where dealing now with *d* electrons strongly hybridized with the surrounding ligands, despite from the Correspondance Principle still being able to reach correct state counting through the designation $1Fe_{II}$ plus $2Fe_{III}$ (or 16 occupied $dp\pi/\pi^*$ and $\sigma^*$ states in total), this definitely cannot now be translated into $d^6 + 2d^5$ of the hard ionic limit $Fe^{2+} + 2Fe^{3+}$. Indeed when, as with inverse spinel structured $Fe_3O_4$, the mixed valence occurs on the same crystallographic site {here the B site - *i.e.* $Fe_{III}(Fe_{II}/Fe_{III})O_4$}, the level of ('static') charge differentiation and ordering at those sites can be much less than integral [161]. The above B site designation as $Fe_{II}/Fe_{III}$ registers in the main that the A site is trivalent. Any further import depends on just what degree of charge and spin ordering arises over the B sites and of the associated lattice response. Where rather slight one may in appropriate circumstances start to speak of a PLD/CDW, as in 2H-$TaS_2$, for which the label $Ta_VTa_{III}$ is neither appropriate nor useful. The 9 Ta atoms per cell become distributed there over several site types [122,123]. In contrast writing $BaBiO_3$ as $Bi_{III}/Bi_V$ has some crystallographic justification [162], although less than does the representation of TlS as $Tl_I/Tl_{III}$, since in the latter case two quite distinct crystal sites are involved of large size difference [163]. In both cases this 'disproportionation' is sufficiently marked for a semiconductor to result, a situation observed too in the T.M. compound $PtI_3$, structurally a 50:50 mix of $Pt_{II}$ and $Pt_{IV}$ [164].

With this lengthy introduction let us now turn to examine the situation that exists in mixed-valent $YBa_2Cu_3O_6$ and $YBa_2Cu_3O_7$. In these two cuprates simple state counting would support $2Cu_{II} + 1Cu_I$ in the former material and $2Cu_{II} + 1Cu_{III}$ in the latter. The increase in oxygen stoichiometry occurs here within the chain region of the structure, and the adopted crystal geometry of a linear two-fold coordination for the chain Cu site in $YBCO_6$ (as in $Cu_2O$) and a square-planar four-fold coordination in $YBCO_7$ (as in $KCuO_2$) makes labelling the *in*-plane sites as $Cu_{II}$ the natural option.

One has to mark here however that $YBa_2Cu_3O_7$ is (i) metallic and (ii) carries a CDW in its chains, in consequence of there being some net charge transfer from the planes into the chain amounting to approximately a sixth of an electron per planar site [1d,165]. Whatever the time averaged site configurations are in the planes and chains of $YBCO_7$ they most certainly are not $d^9$ and $d^8$, neither in parentage nor number.

Now what has been attempted by Temmerman and coworkers in [159] is to address in band structural terms what the prevailing configurational conditions might be, not only in $YBCO_6$ and $YBCO_7$, but also for $YBCO_{6.5}$ (nominally entirely $Cu_{II}$, although note in fact a metal and a superconductor). This closer specification is attempted through comparison against experiment of a whole set of local spin density (LSD) LMTO-ASA band structural and total energy calculations, these implemented using a (*partial*) self-interaction correction (SIC) technique. It is in the manner of implementation of this crucial correction and in the unfortunate terminology employed that the misleading confusion spoken of earlier arises. Much of this originates from the *f*-electron work referred to above. These density functional calculations generate the total energy of the system for a cluster of approximately 100 atoms, sufficiently large to incorporate an AF basal cell doubling if demanded. The SIC correction treatment is extended to a varying selection of Cu orbitals and sites and from this choice we can then identify the one with the least total energy. In the calculation the remaining balance of the states from the widely drawn basis set are treated in standard band structural fashion, both groups of states being embraced within the LSD framework. In addition to appropriate empty spheres a non-spherical Madelung potential array is included. All sites acquire a frozen electron population, this being distributed homogeneously over like sites. Temporal and spatial fluctuations are not examined. What is determined are which orbitals and sites (Cu plane and chain) are best treated within SIC and which not. This is where the wording problems appear. The authors choose to term any SIC $d^8$ site as 'trivalent' and any SIC $d^9$ site as 'divalent' (and indeed then label them as $Cu^{3+}$ and $Cu^{2+}$). Additionally SIC configurations are marked with an asterisk whenever an $x^2$-$y^2$ ($e_g$) state is treated outside the SIC set, and without asterisk when conversely a $z^2$ ($e_g$) state is placed outside. $d^8$ thus specifies that both the up and down spin states ($pd\sigma$ and $pd\sigma^*$) of a particular symmetry are taken to be outside the SIC set, while $d^9$ means only the upper $pd\sigma^*$ state is taken to be outside. I have found it far easier to track what is going on by labelling the delocalized *non*-SIC states, using *h* for $d_{x2-y2}$ and *v* for $d_{z2}$ geometries respectively. In this fashion the corresponding notations in the two schemes for the determined ground-state conditions of YBCO6, 6.5 and 7 (from the range of alternatives considered) are in turn

| | non-SIC states | | SIC states | |
|---|---|---|---|---|
| | plane | chain | plane | chain |
| $YBCO_{6.0}$ | *h* | *vv* | $d^9$ | $d^{8*}$ |
| $YBCO_{6.5}$ | *hh* | *vv* | $d^8$ | $d^{8*}$ |
| $YBCO_{7.0}$ | *hh* | *v* | $d^8$ | $d^{9*}$ |

We see here in metallic $YBCO_{6.5}$ and $YBCO_7$ that both planar $d_{x2-y2}$ orbitals appear best treated (within the present framework) as delocalized, while in Mott-insulating $YBCO_6$ only the antibonding partner would appear extended. Note the latter circumstance is adjudged not to generate metallic behaviour in consequence of the basal AF order supported ($\mu \sim 0.6\ \mu_B$: experimentally 0.45 $\mu_B$ in a larger cell). For $YBCO_7$ the energy difference between the above evaluated ground state condition (*hh*/*v*) and the condition *hh*/*vv*, where both σ and σ* $d_{z2}$ chain states have been placed outside the SIC set is only 0.07 eV, a negligible amount at this level of operation. That should be contrasted with the 0.5 eV in $YBCO_6$ from the ground state (of *h*/*vv*: SIC $d^9/d^{8*}$) up to the lowest energy modification (*hh*/*vv*: SIC $d^{8*}/d^{8*}$). For $YBCO_{6.5}$ the energy sequencing between the latter two state energies becomes inverted and of magnitude 1.0 eV.

Perhaps the most interesting change reported in the paper is the one for $YBCO_6$ upon incorporating at the chain sites all ten *d*-based states into the SIC set, as would be encountered with an atomic-core-like $Cu^{1+}$ condition. The latter is evaluated to possess a state energy 3.9 eV higher than the ground state condition tabulated above. What is more when the corresponding DOS is examined (fig. 1a in [159]) the $d^{10}$ chain state unit exhibits a semi-core status lying some 7.5 eV below $E_F$. With this shell collapse a band gap of 1.6 eV around $E_F$ becomes cleared of states, as compared to a gap of just 0.3 eV for the ground state choice. Unfortunately the paper then proceeds to claim that the chain site in $YBCO_6$ cannot be monovalent. What really is meant here is that it is not $Cu^{1+}$. However it is Cu(I)-like, in that the experimentally favoured condition (as in table above) is very much like that found in $Cu_2O$ and the Cu monohalides. Both calculation and experiment reveal that for these compounds the uppermost V.B. still has strong *d*-state admixing – the *d*-states at monovalence (Madelung potential) have not yet become pulled down into the semi-core [166]. The latter *is* experienced with divalent Zn compounds at shell closure [167], and especially with trivalent Ga [168]. As was represented within figure 3 of my original paper on HTSC [1a], the augmentation in Madelung potential driving down *d*-shell radii is what forces the closed $3d^{10}$ set steeply away from $E_F$. The Madelung potentials are dictated by the degree to which the outer *s* and *p* electrons are being stripped off the various cations under the given stoichiometry and electronegativity of the coordinating ligands. Between $YBCO_6$ and $YBCO_7$ the alternative SIC conditions considered in [159] all become closer in energy in the latter compound in consequence of the better screening in the more metallic material.

It has to be emphasized once more that the various states considered in the above calculations are for static site loadings, homogeneously distributed over similar crystallographic sites, and they do not involve ‘slow’ temporal and spatial charge fluctuation of the type figuring in earlier discussion. [159], despite what is claimed there, is not dealing with valence fluctuations, nor indeed with valency at all in the manner suggested. It is dealing with configuration modification and more specifically with that subset of states in the on-site configuration to be selectively treated by the SIC routine. This said, the paper is of appreciable worth in that it opens up discussion at the local level, as will in due course become necessary when addressing site inhomogeneous behaviour (stripes, etc) and true fluctuational site-loading change, negative-*U* double-loading, etc., all within a self-consistent cluster format.

A final feature which the paper draws attention to is how important local structural relaxation is in association with (SIC) configuration change. Besides the general differences dictated by the changing count of antibonding electrons involved locally, there is as well the strong coordination unit shape change set by the presence or otherwise of the quasi-localized $d^9$ occupancy. This configuration still is strongly J-T active in oxides. Note the planar Cu-O pyramid's apical elongation reduces from 2.450 Å in $YBCO_6$ to 2.389 Å in $YBCO_{6.5}$ to 2.304 Å in $YBCO_7$ as delocalization progresses. Highly delocalized $CuS_2$ displays no structural J-T behaviour at all, becoming isomorphous with $MnS_2$ and $ZnS_2$. By contrast the structure of Mott-insulating CuO is rendered unique through the action of the Jahn-Teller effect [169].

## 5. Closing remarks.

New experimental data from all sectors of HTSC phenomena continue to support the chemical negative-*U* scenario consistently advocated throughout references [1,4,6,7,20,etc.]. Several of those papers were based around a 12-point synopsis of the cuprate circumstance which will provide a useful framework (in somewhat shuffled arrangement here) for summarizing the present work.

**i)** Inhomogeneity associated with mixed-valent nonstoichiometry.

It is on this factor that the two-subsystem nature of the HTSC cuprates is founded. It has been long in evidence from structural probes like EXAFS and PDF [9], and it is good now to see it coming to the forefront in electronic and magnetic probing. The low temperature spatial and energy spectroscopies by STM recently perfected by Pan, Davis and coworkers are an excellent example [74,146,170]. It is good also to see recent NMR/NQR work [171] more receptive to this intrinsic inhomogeneity, first detected by Yoshimura *et al* [147].

**ii)** Anderson disorder localisation.

Systems as close to Mott localization as the present suffer strongly from disorder effects. Hussey *et al*'s low temperature thermal conductivity results on YBCO-124 [135] demonstrate that the disorder in this system is charge led: when time averaged this system is a perfectly regular structure, but that is not so on a short time scale, as revealed through PDF [139].

**iii)** 'Metallic' conduction.

The steady shedding of standard metallicity as the temperature or probe energy is increased is closely described by the Marginal Fermi Liquid theory of Varma, Abrahams and coworkers. However their new treatment [134] attributes the strong scattering involved to 'impurities', while we see the local disordering as intrinsic. The chronic scattering is directly related to the e-e and e-b scattering, and central to the negative-*U* HTSC scenario. This scattering ultimately forces the resistivity beyond the Mott-Ioffe-Regel limit without the customary saturation of e-ph scattering in poor metals [172].

**iv)** Non-rigid-band doping.

The pseudogap behaviour evident in all HTSC properties above $T_c$ manifests itself correspondingly in much theoretical work. The ultra-sensitive behaviour revealed in the band-structural work of Temmerman *et al* [159], as regards the precise choice of SIC-treated states, indicates the limits to which all treatments must be pushed to retain these systems within a band structural framework. That the HTSC systems are universally *p*-type in electrical behaviour is a mark of the depth of the pseudogap. Most HTSC models talk of hole pairing. The present negative-*U* model emphasizes very much that it is electrons which pair.

**v)** Low-dimensionality.

The level of electrical anisotropy which this introduces may in certain cases, as in the intercalated systems [173, 65b], be extreme, yet HTSC persists little affected. The intraplane nature of the HTSC coupling marks its local character. That argument however cannot be reversed, for no 3D mixed-valent cuprates are superconducting; instead these show magnetic behaviour. The specifically 2-dimensional structuring introduces other important consequences to be noted below.

**vi)** Fermi Surface geometry.

The simple crystal structure leads to a simple Fermi surface. Photoemission finally has been able to detect that a two-layer crystal structure in Bi-2212 indeed leads to a two-sheeted Fermi surface [174]. The HTSC systems are by no means so electronically perturbed that all *k*-space physics is lost. ARPES not only has tracked the F.S. geometry which is dictated by the square-planar cuprate structures, but recorded also how the density of states is modified around that F.S. by the onset of superconductivity. The recent experiments of Valla *et al* [47-49] among others reveal additionally how the self-energy of the 'quasi particles' is modified around F.S., with scattering activity dominated by what goes on in proximity to the M-point DOS saddles. This is the action securing negative-*U*, shell-closure pairing in the present scenario.

**vii)** Density wave states.

With a crystal structure and Fermi surface that is low-dimensional and as simple as the present, it is remarkable the materials opt for a superconductive rather than some density wave ground state. It is observed that the chain states in non-superconducting $PBCO_7$ do in fact support a CDW below 120 K [175]. Possibly the HTSC systems are too disturbed by their strong electronic scattering to settle normally into such a state, and the prevailing non-stoichiometric electron count current in the planes is probably disruptive too. Fortunately moreover the planar F.S. does not show particularly good nesting geometry. Stripes, of course, are a form of density wave, but this phenomenon we do not regard as Fermi surface driven, but on the contrary as one impressing a geometry determined solely by the overall electron count and electronic phase separation [1d,e].

**viii)** Jahn-Teller effect.

The stripes are an expression of dopant charge self-organization, and they come to define the format of the inherent two-subsystem structuring. That organization is without a doubt strongly assisted by the Jahn-Teller apical elongation of coordination units at sites carrying a $d^9$ electron count [1e]. Those sites are in the main the ones in the regions bounded by the stripes themselves, forming there the lower valent majority subsystem.

**ix)** Secondary distortions.

In perovskite oxides like the present, additional distortions governed by ion size accommodation are very common, eg. the $GdFeO_3$ orthorhombic distortion. The LTO and LTT low temperature structures of LSCO are of this form, showing mutual coordination unit tilting. This is compounded by the fact that in many HTSC systems such as YBCO even the basic coordination units are not centrosymmetric, so the cations never sit at a special point but are free to shift somewhat as a function of temperature. It ought to be borne in mind that the oxygen and copper atoms in YBCO are not indeed strictly coplanar. Much but not all of this detail appears peripheral to the HTSC outcome, though it often becomes registered in the phonon responses and in isotopic shifts [70,71,105].

**x)** Reduced magnetic moments.

The magnetic moments in the HTSC systems fortunately are low. Square-planar $d^8$ customarily takes a low spin S = 0 arrangement, as in $KCuO_2$ or $PdCl_2$. Quantum effects furthermore hold down the S = ½ moment of $d^9$, as in pure $La_2CuO_4$, to a value of around 0.6 $\mu_B$ [176], and the loss of moment is promoted by any advance in delocalization brought about with *p*/*d* hybridization or nonstoichiometry and metallization. There has been considerable speculation recently about the very small apparent moments detected by neutron diffraction in $YBCO_7$ [177], in connection with a postulated existence of orbital currents in these systems [178]. However just released μSR work [179] serves to affirm that the detected effects emerge not from a uniform distribution of very small moments, but from an inhomogeneous distribution of somewhat larger moments. This befits the two-subsystem arrangement, in which certain $d^9$ atoms are sited in special locations with regard to the stripes. In [1e] we indicated that the likely origin of such 'revealed' moments is spin canting.

**xi)** RVB vs antiferromagnetic ordering.

Neutron spin diffraction into incommensurately arranged spots was treated in [1d,e] as referring to such stripe-driven spin organization. By contrast the strong $\pi\pi$ commensurate neutron diffraction seen only near and below $T_c$ was understood as relating not to antiferromagnetic ordering (even of SRO form), but to the spin-flip pair breaking of local pairs. The two halves of the pair are returned to neighbouring coordination units, preferentially with a z-axis component as well. Likewise local pairs are seen as being formed in joint $\pi\pi$ scatterings from neighbour *k*-space M-point saddles into the maximally antibonding corners of the Brillouin zone [1f]. Antiferromagnetic coupling in these highly *p*/*d* hybridized systems displaying appreciable delocalization is replaced within the given geometry by RVB square-plaquet coupling over the majority subsystem. RVB theory continues to be enhanced and applied to new S = ½ systems [180].

**xii)** Disproportionation.

Disproportionation is to be avoided. AgO disproportionates as Ag(I)Ag(III) and in consequence is non-metallic. What disproportionation usually is founded upon is the attainment of some stable local electronic configuration: in AgO full shell $d^{10}$ Ag(I) and square-planar $d^8$ Ag(III). This is the same option that propels disproportionation in perovskite $CsAuCl_3$ at low temperature, eliminating thereby the metallicity of high temperatures [162c]. The presently advocated model makes use of the high stability of the full shell electron count attained in its double-loading negative-*U* fluctuations. Because the latter are attained at the sites of the minority subsystem of raised valence they are particularly effective in countering the positive-*U* Hubbard term, in the current tight-binding circumstances of 3 to 4 eV [see 1f, fig 1].

In the present paper we have seen how the laser based optical and electric work of Li [99], Little [97], Demsar [98] and their coworkers strongly support a negative-*U* value of the desired magnitude of –3 eV per pair or –1.5 eV per partner, so as to counter the above positive term and leave the bosons effectively degenerate with the Fermi energy. This permits electrons and bosons to interact strongly and marks the most favourable 'intermediate' circumstance for maximizing $T_c$, rather than being in the extreme BCS or hard boson limits [181,182]. The work of Pickett and coworkers [104] has shown how the lattice is strongly coupled into the support of the HTSC outcome, and Little *et al* [97] point to the same conclusion. The work of Corson *et al* [88,92] is seen as supporting the same understanding reached earlier from optical work by Homes *et al* [183] that the dynamic flux between the single and pair electron condition leads to a very substantial number of bosons being left at any one instant outside the coherent superconducting bosonic condensate. It is these bosons residing outside the $k = 0$ condensate which have been taken in the present scenario to be responsible for the low energy little dispersing collective mode that interacts so strongly with the quasiparticles in the vicinity of the vHs saddles and from where the pairs are drawn. As analyzed by Eschrig and Norman [125] such interaction and self-energy modification of the quasi-particles goes on to be responsible for the characteristic 'resonant' form to the ARPES spectra at these *k*-values.

A good many theorists over the years have abstractly developed negative-*U* models of considerable potential for application to HTSC, but none has felt drawn to proceed into the complexities set out above which represent the actual circumstances prevailing in the real mixed-valent cuprate systems. One might single out the original work of Micrias, Ranninger and Robaszkiewicz [184] and its recent developments, especially that by Domanski [185]. Likewise the early work of Friedberg and Lee [186] became very promising in the developments secured by Blaer, Ren, Tchernyshyov and others [187]. The two-subsystem negative-*U* treatment suggested by Stocks et al in 1991 [182], combining their local pair Hubbard approach with the CPA alloy formulism never fully materialised. I would like to encourage all these workers and others now treading the negative-*U* path, such as Casas, Tolmachev and and their many coworkers [188], to adopt some of the present empirically founded proposals into their formal analysis.

The situation as presented clearly is many-sided and complex, but that is what is to be expected with HTSC so uniquely rare a phenomenon. We have noted how HTSC in the cuprates needs the local tight-binding circumstance. It occurs in a nanostructural two-subsystem milieu. Strong *p*/*d* hybridization supports a low pair-breaking situation in promoting RVB within the $d^9$ subsystem. The 2D nature of the crystalline and *k*-space properties encourages a strong lattice response to electronic order and long-lived fluctuations. The square-planar Cu-O local coordination is crucial in establishing a F.S. geometry with strong saddle-point features just below $E_F$ from which heavy shell-filling metastable pairings can be readily constructed. All these properties are to some degree or other dictated by the particular position of copper within the periodic table, right at the end of the first T.M. series. This is what establishes the uniqueness of what is displayed by Hg-1223 and its less spectacular partners, and what still to the author makes the epithet 'chemical' most appropriate.

**References**

[1] *a* Wilson J A, 1988 *J. Phys. C: Sol. St. Phys*. **21** 2067-2102;

*b* Wilson J A, 1989 *Int. J. Mod. Phys*. B **3** 691-710.

*c* Wilson J A, 1994 *Physica C* **233** 332-348.

*d* Wilson J A and Zahrir A, 1997 *Rep. Prog. Phys*. **60** 941-1024.

*e* Wilson J A, 2000 *J. Phys.: Condens. Matter* **12** 303-310.

*f* Wilson J A, 2000 *J. Phys.: Condens. Matter* **12** R517-R547.

[2] Shirane G, Birgeneau R J, Endoh Y, Gehring P, Kastner M A, Kitazawa K, Kojima H, Tanaka I, Thurston T R and Yamada K, 1989 *Phys. Rev. Lett.* **63** 330.

Mason T E, Aeppli G, Hayden S M, Ramirez A P and Mook H A, 1993 *Phys. Rev. Lett*. **71** 919.

Tranquada J M, Sternlieb B J, Axe J D, Nakamura Y and Uchida S, 1995 *Nature* **375** 561.

[3] McKernan S, Steeds J W and Wilson J A, 1982 *Physica Scripta*, **T1**, 74.

Fung K K, McKernan S, Steeds J W and Wilson J A, 1981 *J.Phys.C:Sol.St.Phys.***14** 5417.

Wilson J A, 1985 *J. Phys. C: Sol. St. Phys*. **15** 591;

1990 *J. Phys.: Condens. Matter* **2** 1683.

[4] Wilson J A, 1998 *J. Phys.: Condens. Matter* **10** 3387-3410.

Farbod M, Giblin S, Bennett M and Wilson J A, 2000 *J. Phys.: Condens. Matter* **12** 2043-2052.

[5] Martinez G, Mignot J M, Chouteau G, Chang K J, Dacorogna M M and Cohen M L, 1986 *Physica Scripta*, T**13** 226.

Valyanskaya T V and Stepanov G N, 1993 *Solid State Commun*. **86** 723. {Si}

Kawamura H, Shirotani I, and Tachikawa K, 1984 *Solid State Commun*. **49** 879. {P}

Struzkin V V, Hemley R J, Mao H K and Timofeev Y A, 1998 *Nature* **390** 382. {S}

[6] Rosseinsky M J, Ramirez A P, Glamm S H, Murphy D W, Haddon R C, Hebard A F, Palstra T T M, Kortan A R, Zahurak S M and Makija A V, 1991 *Phys. Rev. Lett.* **66** 2830.

Wilson J A, 1991 *Physica* C **182** 1-10.

[7] Yamanaka S, Hotehama K and Kawaji H, 1998 *Nature* **392** 580.

Wilson J A , 1999 *Supercond. Sci. Technol*. **12** 649-653.

Tou H, Maniwa Y, Koiwasaki T and Yamanaka S, 2000 *Phys. Rev.* B **63** 020508.

[8] Nagamatsu J, Nakagawa N, Muranaka T, Zenutani Y and Akimitsu J, 2001 *Nature* **410** 63.

Bud'ko S L, Lapertot G, Petrovic C, Cunningham C E, Anderson N and Canfield P C, 2001 *Phys. Rev. Lett.* **86** 1877.

Finnemore D K, Ostenson J E, Bod'ko S L, Lapertot G and Canfield P C, 2001 *Phys. Rev. Lett.* **86** 2420.

[9] Egami T and Billinge S J L, p.265*ff*. in *Physical Properties of High Temperature*

*Superconductors,* Ed: D.Ginsburg. (Pub: World Scientific, Singapore, 1996).
Billinge S J L, Bozin E S, Gutmann M and Takagi H, 2000 *J. Supercond.* **13** 713.
Oyanagi H, Saini N L and Bianconi A, 2000 *Int. J. Mod. Phys.* B **14** 3623.
[10] Grimaldi C, Cappelluti E, Pietronero L and Strässler S, 2000 *Int. J. Mod. Phys*. B **14** 2950.
Botti M, Cappelluti E, Grimaldi C and Pietronero L, 2000 *Int. J. Mod. Phys.* B **14** 2976.
Cappelluti E, Grimaldi C, Pietronero L, Strässler S and Ummarino G A, 2001 *Eur. Phys. J.* B **21** 383.
[11] Alexandrov A S, 1999 *Physica* C **316** 239.
Alexandrov A S and Kabanov V V, 1999 *Phys. Rev.* B **59** 13628.
Alexandrov A S, *arXiv:cond-mat*/0102189.
[12] Pringle D J, Williams G V M and Tallon J L, 2000 *Phys. Rev.* B **62** 12527.
[13] Zhang Y, Ong N P, Anderson P W, Bonn D A, Liang R and Hardy W N, 2001 *Phys. Rev. Lett.* **86** 890.
[14] Misochko O V and Sherman E Ya, 1998 *Int. J. Mod. Phys.* **12** 2455. Cardona M, 1999 *Physica* C **317**/**8** 30.
Limonov M F, Tajima S and Yamanaka A, 2000 *Phys. Rev.* B **62** 11859.
Sacuto A, Caysol J, Monod P and Colson D, 2000 *Phys.Rev.* B **61** 7122.
[15] Schluter M A, Lannoo M, Needels M F, Baraff G A, 1992 *Phys. Rev. Lett.* **69** 21.
[16] Kortus J, Mazin I I, Belashchenko K D, Antropov V P and Boyer L L, *arXiv:cond-mat*/0101446.
Lorenz B, Meng R L and Chu C W, *arXiv:cond-mat*/0102264.
N.B. very specific coupling to ΓA holes in B-B basal net that is evident in Bohnen K-P, Heid R and Renker B, *arXiv:cond-mat*/0103319.
Similarly now see Liu A Y, Mazin I I and Kortus J, *arXiv:cond-mat* /0103570.
[17] Sun Y P, Song W H, Dai J M, Zhao B, Du J J, Wen H H and Zhao Z X, *arXiv:cond-mat*/0103101. (mixed phase sample)
Schön J H, Kloc C and Batlogg B, 2000 *Nature* **408** 549.
[18] Allen P B and Mitrovic B, 1982 *Solid State Physics,* p.371ff, Eds: Ehrenreich H, Seitz F and Turnbull D. (Pub: Academic, New York).
Evans R, Gaspari G D and Gyorffy B L, 1973 *J. Phys. F; Met. Phys.* **3** 39.
N.B. the excessively small $\mu^*$ needed to accommodate the new $MgB_2$ data to the Allen-Dynes-McMillan formulation: see Kong Y, Dolgov O V, Jepson O and Andersen O K, *arXiv:cond-mat*/0102499.
[19] Goodrich R G, Grienier C, Hall D, Lacerda A, Haanappel E G, Rickel D, Northington T, Schwarz R, Mueller F M, Koelling D D, Vuillemin J, van Bockstal L, Norton M L and Lowndes D H, 1993 *J. Phys. Chem. Solids* **54** 1251.
This high field work was at the limit of detection and figures 5 and 6 do not look compatible.
[20] Wilson J A and Farbod M, 2000 *Supercond. Sci.* Technol. **13** 307-322.
[21] Aubin H, Behnia K, Ribault M, Gagnon R and Taillefer L,

1997 *Phys. Rev. Lett.* **78** 2624; 1997 *Z. Phys.* B **104** 175.
Ausloos M and Houssa M, 1999 *Supercond. Sci. Technol.* **12** 103.
[22] McGuire J J, Windt M, Startseva T, Timusk T, Colson D and Viallet-Guillen V, 2000 *Phys. Rev.* B **62** 8711.
Timusk T and Statt B, 1999 *Rep. Prog. Phys.* **62** 61-122.
[23] Loram J W, Mirza K A, and Cooper J R, pp 77 - 97 in *Research Review 1998 HTSC*. [Ed: Liang W Y; Pub: IRC, Univ. of Cambridge, 1998].
Loram J W, Luo J, Cooper J R, Liang W Y and Tallon J L, 2000 *Physica* C**341-8** 831, 2001 *J. Phys. Chem. Solids* **62** 59.
[24] Fujita S, Obata T, Morabito D L and Shane T F, 2001 *Phys. Rev.* B **63** 054402.
[25] Carretta P, Lascialfri A, Rigamonti A, Rosso A and Varlamov A, 2000 *Phys. Rev.* B **61** 12420.
Gorny K R, Vyaselev O M, Pennington C H, Hammel P C, Hults W L, Smith J L, Baumgartner J, Lemberger T R, Klamut P and Dabrowski B, 2001 *Phys. Rev.* B **63** 064513.
[26] Yu R C, Naughton M J, Yan X, Chaikin P M, Holtzberg F, Greene R L, Stuart J and Davies P, 1988 *Phys. Rev.* B **37** 7963.
[27] Harris J M, Yan Y F, Matl P, Ong N P, Anderson P W, Kimura T, and Kitazawa K, 1995 *Phys. Rev. Lett.* **75** 1391.
[28] Dai P, Yethiraj M, Mook H A, Lindemer T B and Dogan F, 1996 *Phys. Rev. Lett.* **77** 5425.
[29] Aeppli G, Mason T E, Hayden S M, Mook H A & Kulda J, 1997 *Science* **278** 1432.
[30] Dai P, Mook H A, Hayden S M, Aeppli G, Perring T G, Hunt R D and Dogan F, 1997 *Science* **278** 1432.
[31] Lake B, Aeppli G, Mason T E, Schröder A, McMorrow D F, Lefmann K, Isshiki M, Nohara M, Takagi H and Hayden S M, 1999 *Nature* **400** 43.
[32] Dai P, Mook H A, Hayden S M, Aeppli G, Perring T G, Hunt R D and Dogan F, 1999 *Science* **284** 1344.
[33] Mook H A, Dai P, Dogan F and Hunt R D, 2000 *Nature* **404** 729.
[34] Dai P, Mook H A, Aeppli G, Hayden S M and Dogan F, 2000 *Nature* **406** 965.
[35] Dai P, Mook H A, Hunt R D and Dogan F, 2001 *Phys. Rev.* B **63** 054525.
[36] Mook H A, Dai P and Dogan F, *arXiv:cond-mat*/0102047.
[37] Fong H F, Bourges P, Sidis Y, Regnault L P, Ivanov A, Gu G D, Koshizuka N and Keimer B, 1999 *Nature* **398** 588.
[38] Fong H F, Bourges P, Sidis Y, Regnault L P, Bossy J, Ivanov A, Milius D L, Aksay I A and Keimer B, 2000 *Phys. Rev.* B **61** 14773.
[39] He H, Sidis Y, Bourges P, Gu G D, Ivanov A, Koshizuka, N, Liang B, Lin C T, Regnault L P, Schoenherr E and B Keimer, 2001 *Phys. Rev. Lett.* **86** 1610.
[40] Ding H, Norman M R, Campzano J C, Randeria M, Bellman A F, Yokoya T, Takahashi T, Mochiku T, Kadowaki K, 1996 *Phys. Rev.* B **54** R9678.

[41] Norman M R, Ding H, Campuzano J C, Takeuchi T, Randeria M, Yokoya T, Takahashi T, Mochiku T and Kadowaki K, 1997 *Phys. Rev. Lett.* **79** 3506.

Norman M R and Ding H, 1998 *Phys. Rev.* B **57** 11089.

Norman MR, Randeria M, Ding H and Campuzano J C, 1998 *Phys. Rev.* B **57** R11093.

[42] Norman M R, Ding H, Randeria M, Campuzano J C, Yokoya T, Takeuchi T, Mochiku T, Kadowaki K, Guptasarma P and Hinks D G, 1998 *Nature* **392** 157.

[43] Mesot J, Norman M R, Ding H, Randeria M, Campuzano J C, Paramekanti A, Fretwell H M, Kaminski A, Takeuchi T, Yokoya T, Sato T, Takahashi T, Mochiku T and Kadowaki K, 1999 *Phys. Rev. Lett.* **83** 840.

[44] Kaminski A, Mesot J, Fretwell H, Campuzano J C, Norman M R, Randeria M, Ding H, Sato T, Takahashi T, Mochiku T, Kadowaki K and Hoechst H, 2000 *Phys. Rev. Lett.* **84** 1788.

Mesot J, Kaminski A, Fretwell H M, Rosenkranz S, Campuzano J C, Norman M R, Ding H, Randeria M and Kadowaki K, 2000 *Int. J. Mod. Phys.* B **14** 3596.

[45] Kaminski A, Randeria M, Campuzano J C, Norman M R, Fretwell H, Mesot J, Sato T, Takahashi T and Kadowaki K, 2001 *Phys. Rev. Lett.* **86** 1070.

[46] Mesot J, Boehm M, Norman M R, Randeria M, Metoki N, Kaminski A, Rosenkranz S, Hiess A, Fretwell H M, Campuzano J C and Kadowaki K, *arXiv:cond-mat*/0102339.

[47] Valla T, Federov A V, Johnson P D, Wells B O, Hulbert S L, Li Q, Gu G D and Koshizuka N, 1999 *Science* **285** 2110.

[48] Valla T, Federov A V, Johnson P D, Li Q, Gu G D and Koshizuka N, 2000 *Phys. Rev. Lett.* **85** 828.

[49] Johnson P D, Valla T, Federov A V, Yusof Z, Wells B O, Li Q, Moodenbaugh A R, Gu G D, Koshizuka N, Kendziora C, Jian S and Hinks D G, *arXiv:cond-mat*/0102260.

[50] Bogdanov P V, Lanzara A, Kellar S A, Zhou X Z, Lu E D, Zheng W J, Gu G, Shimoyama J-I, Kishio K, Ikeda H, Yoshizaki R, Hussain Z and Shen Z X, 2000 *Phys. Rev. Lett.* **85** 2581.

[51] Shen Z X, Lanzara A and Nagaosa N, *arXiv:cond-mat*/0102244.

[52] Lanzara A, Bogdanov P V, Zhou X J, Kellar S A, Feng D L, Lu E D, Yoshida T, Eisaki H, Fujimori A, Kishio K, Shimoyama J-I, Noda T, Uchida S, Hussain Z and Shen Z-X, 2001 *Nature* **412** 510.

[53] Feng D L, Armitage N P, Lu D H, Damascelli A, Hu J P, Bogdanov P, Lanzara A, Ronning F, Shen K M, Eisaki H, Kim C, Shen Z-X, Shimoyama J-i and Kishio K, *arXiv;cond-mat*/0102385 v2.

[54] Chuang Y D, Gromko A D, Federov A, Dessau D S, Aiura Y, Oka K, Ando Y, Eisaki H and Uchida S I, *arXiv;cond-mat*/0102386.

[55] Lu D H, Feng D L, Armitage N P, Shen K M, Damascelli A, Kim C, Ronning F, Shen Z-X, Bonn D A, Liang R, Hardy W N, Rykov A I and Tajima S, 2001 *Phys. Rev. Lett.* **86** 4370.

[56] Fujimori A, Ino A, Yoshida T, Mizokawa T, Nakamura M, Kim C, Shen Z-X, Kishio K, Kakeshita T, Eisaki H and Uchida S, 2001 *J. Phys. Chem. Solids* **62** 15.

[57] Tohyama T and Maekawa S, 2000 *Supercond. Sci. Technol.* **13** R17.

[58] Manzke R, Müller R, Janowitz C, Schneider M, Krapf A and Dwelk H, 2001 *Phys. Rev.* B **63** 100504(R).

[59] Wilson J A, 1972 *Adv. in Phys.* **21** 143 - 198.
1984 *NATO ASW* B **113** 657-758;
ed: P Phariseau and W M Temmerman, (New York: Plenum).
1985 Ch.9 (pp.215-260) in '*The Metallic and Non-metallic States of Matter',*
entitled 'The Mott transition for binary compounds, including a case study on $Ni(S_{1-x}Se_x)_2$'.

[60] Mizuno Y, Tohyama T and Maekawa S, 1998 *Phys. Rev.* B **58** R14713.

[61] Kageyama H, Onizuka K, Ueda Y, Nohara M, Suzuki H and Takagi H, 2000 *JETP* **40** 129.

[62] Itoh T, Fueki K, Tanaka Y and Ihara H, 1999 *J. Phys. Chem. Sol.* **60** 41.

[63] Kontos A G, Han Z P and Dupree R, 1999 *Physica* C **317/8** 569.
Tokunaga Y, Ishida K, Yoshida K, Mito T, Kitaoka Y, Nakayama Y, Shimoyama J, Kishio K, Narikiyo O and Miyake K, 2000 *Physica* B **284/8** 663.

[64] Suzuki M and Watanabe T, 2000 *Phys. Rev. Lett.* **85** 4787.

[65] Krasnov V M, Kovalev A E, Yurgens A and Winkler D, 2001 *Phys. Rev. Lett.* **86** 2657.
Krasnov V N, Yurgens A, Winkler D, Delsing P and Claeson T, 2001 *Physica* C **352** 89.

[66] Uchida S, Tamasaku K and Tajima S, 1996 *Phys. Rev.* B **53** 14558.

[67] Mosqueira J, Carballeira C, Ramallo M V, Torrón C, Veira J A and Vidal F, 2001 *Europhys. Lett.* **53** 632.

[68] Carballeira C, Currás S R, Viña J, Veira J A, Ramallo M V and Vidal F, 2001 *Phys. Rev*. B **63** 144515.

[69] Bennett M, 1999 PhD thesis, University of Bristol.

[70] Rubio Temprano D, Mesot J, Janssen S, Conder K, Furrer A, Sokolov A, Trounov V, Kazakov S M, Karpinski J and Müller K A, 2001 *Eur. Phys. J.* B **19** 5.

[71] Hofer J, Conder K, Sasagawa T, Zhao G-M, Schneider T, Karpinski J, Willemin M, Keller H and Kishio K, 2000 *J. Supercond.* **13** 963.

[72] Meingast C, Pasler V, Nagel P, Rykov A, Tajima S and Olsson P, 2001 *Phys. Rev. Lett.* **86** 1606.

[73] Torrance J B, Bezinge A, Nazzal A I, Huang T C, Parkin S S P, Keane D T, La Placa S J, Horn P M and Held G A, 1989 *Phys. Rev.* B **40** 8872.

[74] Hudson E W, Pan S H, Gupta A K, Ng K-W and Davis J C, 1999 *Science* **285** 88.

[75] Bernhard C, Tallon J L, Blasius Th, Golnik A and Niedermayer C, 2001 *Phys. Rev. Lett.* **86** 1614.

[76] Uemura Y J, 2000 *Int. J. Mod. Phys.* B **14** 3703.

[77] Panagopoulos C, Rainford B D, Cooper J R, Lo W, Tallon J L, Loram J W, Betouras J, Wang Y S and Chu C W, 1999 *Phys. Rev.* B **60** 14617.

[78] Nakano T, Momono N, Matsuzaki T, Nagata T, Yokoyama M, Oda M and Ido M,
1999 *Physica* C**317/8** 575.

[79] Ye J, Sadewasser S, Schilling J S, Zou Z, Matsushita A and Matsumoto T,
1999 *Physica* C **328** 111.
Staub U, Shi M, O'Conner A G, Kramer M J and Knapp M, 2001 *Phys. Rev.* B **63** 134522.

[80] Akoshima M, Koike Y, Watanabe I and Nagamine K,
2000 *Phys. Rev.* B **62** 6761 . {YBCO}
Watanabe I, Akoshima M, Koike Y, Ohira S and Nagamine K,
2000 *Phys. Rev.* B **62** 14524. {BSCCO}

[81] Mahajan A V, Alloul H, Collin G and Marucco J F, 2000 *Eur. Phys. J.* B **13** 457.

[82] Teitelbaum G B, Abu-Shiekah I M, Bakharev O, Brom H B and Zaanen J,
2000 *Phys. Rev.* B **63** 020507.

[83] Katano S, Sato M, Yamada K, Suzuki T and Fukase T, 2000 *Phys. Rev.* B **62** R14677.

[84] Ratcliffe J W, Loram J W, Wade J M, Witschek G and Tallon J L,
1996 *J. Low Temp. Phys.* **105** 903.

[85] Nachumi B, Keren A, Kojima K, Larkin M, Luke G M, Merrin J, Tchernyshyov O,
Uemura Y J , Ichikawa N, Goto M and Uchida S, 1996 *Phys. Rev. Lett.* **77** 5421.

[86] Maggio-Aprile I, Renner C, Erb A, Walker E and Fischer Ø,
1995 *Phys. Rev. Lett.* **75** 2754.

[87] Ramallo M V, Pomar A and Vidal F, 1996 *Phys. Rev.* B **54** 4341.

[88] Corson J, Mallozi R, Orenstein J, Eckstein J N and Bozovic I, 1999 *Nature* **398** 221,
and accompanying comment by Millis A J, *ibid.* p.193.

[89] Berezinskii V L, 1972 *Sov. Phys. JETP* **34** 610**.**
Kosterlitz J M and Thouless D J, 1973 *J. Phys. C: Solid State Phys.* **6** 1181.

[90] Tallon J L, 1998 *Phys. Rev.* B **58** R5956.
The carrier contents here are extracted by employing the parabola rule, *ibid* (1995) B **51** 12911.
Note should be paid to how complicated the real doping stoiciometry is in 'BSCCO-2212'; see
Jean F, Collin G, Andrieux M, Blanchard N and Marucco J-F, 2000 *Physica* C **339** 269.
We encoutered a similar complex circumstance in HBCO-1201 [4b].

[91] Dordevic S V, Singley E J, Basov D N, Kim J H, Maple M B and Bucher E,
*arXiv:cond-mat*/0102455.

[92] Corson J, Orenstein J, Oh S, O'Donnell J and Eckstein J N,
2000 *Phys. Rev. Lett.* **85** 2569.

[93] Varma C M, 1989 *Int. J. Mod. Phys.* B **3** 2083.
Varma C M, Littlewood P, Schmitt-Rink S, Abrahams E and Ruckenstein A E,
1989 *Phys. Rev. Lett.* **63** 1996.

[94] Tilley D R and Tilley J, *Superfluidity and Superconductivity*, 2nd edn. (Hilger, Bristol 1986).

[95] Holcomb M J, Perry C l, Collman J P and Little W A, 1996 *Phys. Rev.* B **53** 6734.

[96] Stevens C J, Smith D, Chen C, Ryan J F, Pobodnik B, Mihailovic D, Wagner G A and

Evetts J E, 1997 *Phys. Rev. Lett.* **78** 2212;
and comment Mazin I I, 1998 *Phys. Rev. Lett.* **80** 3664,
with reply Stevens C J *et al*, *ibid* 3665.

[97] Little W A and Holcomb M J, 2000 *J. Supercond.* **13** 695.
Little W A, Collins K and Holcomb M J, 1999 *J. Supercond.* **12** 89.

[98] *a* Kabanov V V, Demsar J, Pobodnik B and Mihailovic D, 1999 *Phys. Rev.* B **59** 1497.
*b* Demsar J, Podobnik B, Kabanov V V, Wolf Th and Mihailovic D, 1999 *Phys. Rev. Lett.* **82** 4918.
*c* Kabanov V V, Demsar J and Mihailovic D, 2000 *Phys. Rev.* B **61** 1477.
*d* Demsar J, Hudej R, Karpinski J, Kabanov V V and Mihailovic D, 2001 *Phys. Rev.* B **63** 054519.

[99] Li E, Li J J, Sharma R P, Ogale S B, Cao W L, Zhao Y G, Lee C H and Venkatesan T, *arXiv:cond-mat*/0103046.

[100] Genzel L, Wittlin A, Bauer M, Cardona M, Schönherr E and Simon A, 1989 *Phys. Rev.* B **40** 2170.

[101] Homes C C, McConnell A W, Clayman B P, Bonn D A, Liang R, Hardy W N, Inoue M, Negishi H, Fournier P and Greene R L, 2000 *Phys. Rev. Lett.* **84** 5391.

[102] Mostoller M, Zhang J, Rao A M and Ecklund P C, 1990 *Phys. Rev.* B **41** 6488.

[103] Pintschovius L, Pyka N, Reichardt W, Rumiantsev A Yu, Mitrofanov N L, Ivanov A S, Collin C and Bourges P, 1991 *Physica* C **185/9** 156.
Pintschovius L and Reichardt W, p.165 in *Neutron Scattering in Layered Copper-Oxide Superconductors.* Ed: A Furrer. (Vol. 20 of 'Physics and Chemistry of Materials with Low-Dimensional Structures'; Pub: Kluwer Academic, Dordrecht, 1998).

[104] Krakauer H, Pickett W E and Cohen R E, 1993 *Phys. Rev.* B **47** 1002

[105] Franck J P, 1997 *Physica* C **282-7** 198.
Kishore R, pp 23-58 in *Studies in High Temperature Superconductors*, vol. 29. (Ed: A Narlikar; Pub: Nova Science Publ. Inc., Commack N.Y., 1999).

[106] Lanzara A, Zhao G-m, Saini N L, Bianconi A, Conder K, Keller H and Müller K A, 1999 *J. Phys.: Condens. Matter* **11** L541.

[107] Bianconi A, Saini N L, Rosetti T, Lanzara A, Perali A, Missori M, Oyanaga H, Yamaguchi H, Nishihara Y and Ha D H, 1996 *Phys. Rev.* B **54** 12018.
Saini N L, Lanzara A, Oyanagi H, Yamaguchi H, Oka K, Ito T and Bianconi A, 1997 *Phys. Rev.* B **55** 12759.

[108] McQueeney R J, Petrov Y, Egami T, Yethiraj M, Shirane G and Endoh Y, 1999 *Phys. Rev. Lett.* **82** 628.
Pintschovius L and Braden M, 1999 *Phys. Rev.* B **60** R15039

[109] Mook H A and Dogan F, 1999 *Nature* **401** 145.

[110] Tranquada J M, Axe J D, Ichikawa N, Moodenbaugh A R, Nakamura Y and Uchida S, 1997 *Phys. Rev. Lett.* **78** 338.

[111] Mook H A and Dogan F, 2001 *arXiv:cond-mat*/0103037.

[112] Schaerpf O, Capellman H, Brueckel Th, Comberg A and Passing H, 1990 *Zeit. Phys.* B - *Cond. Matter* **78** 345.

Rossat-Mignod J, Regnault L P, Bourges P, Burlet P, Vettier C and Henry J Y, 1993 *Physica* C **192** 109.

[113] Wakimoto S, Birgeneau R J, Lee Y S and Shirane G, 2001 *Phys. Rev.* B **63** 172501.

[114] Mihailovic D and Kabanov V V, 2001 *Phys. Rev.* B **63** 054505.

[115] Nimori S, Sakita S, Nakamura F, Fujita T, Hata H, Ogita N and Udagawa M, 2000 *Phys. Rev.* B **62** 4142.

[116] Gay P, Stevens C J, Smith D C, Ryan J F, Yang G and Abell J S, 1999 *J. Low Temp. Phys.* **117** 1025.

[117] Smith D C, Gay P, Stevens C J, Wang D Z, Wang J H, Ren Z F and Ryan J F, 1999 *J. Low Temp. Phys.* **117** 1057.

[118] Eesley G L, Heremans J, Meyer M S, Doll G L and Liou S H, 1990 *Phys. Rev. Lett.* **65** 3445.

[119] Norman M R, Kaminski A, Mesot J and Campuzano J C, 2001 *Phys. Rev.* B **63** 140508(R).

[120] Valla T, Federov A V, Johnson P D and Hulbert S L, 1999 *Phys. Rev. Lett*. **83** 2085.

[121] Valla T, Federov A V, Johnson P D, Xue J, Smith K E and DiSalvo F J, 2000 *Phys. Rev. Lett.* **85** 4759.

[122] Wilson J A, DiSalvo F J and Mahajan S, 1975 *Adv. in Phys*. **24** 117-201.

[123] Wilson J A, 1977 *Phys. Rev*. B **15** 5748-5757.

Withers R L and Wilson J A, 1986 *J Phys. C; Solid State Phys*. **19** 4809-4845.

[124] McMillan W L, 1975 *Phys. Rev*. B **12** 1197.

[125] Eschrig M and Norman M R, 2000 *Phys. Rev. Lett.* **85** 3261.

[126] Pines D and Monthoux P, 1995 *J. Phys. Chem. Solids* **56** 1651.

[127] Chubukov A V, 1998 *Europhys. Lett.* **44** 655.

Abanov Ar and Chubukov A V, 1999 *Phys. Rev. Lett.* **83** 1652.

Haslinger R, Chubukov A V and Abanov Ar, 2000 *Phys. Rev.* B **63** 020503(R).

Abanov Ar, Chubukov A V and Schmalian J, *arXiv:cond-mat*/0005163, 2001 *Phys. Rev*. B **63** 180510.

[128] Sigrist M, Agterberg D, Furusaki A, Honerkamp C, Ng K K, Rice T M and Zhitomirsky M E, 1999 *Physica* C **317-8** 134.

Nomura T and Yamada K, 2000 *J. Phys. Soc. Jpn.* **69** 1856.

[129] Varma C M, 1999 *Phys. Rev. B* **60** R6973.

[130] Fulde P, Yaresko A N, Zvyagin A A and Grin Y, 2001 *Europhys. Lett.* **54** 779.

Lee S-H, Qiu Y, Broholm C, Ueda Y and Rush JJ, 2001 *Phys. Rev. Lett.* **86** 5554.

[131] Grenier B, Cepas O, Regnault L P, Lorenzo J E, Ziman T, Boucher J P, Hiess A,

Chatterji T, Jegoudez J and Revcolevschi A, 2001 *Phys. Rev. Lett.* **86** 5966.
Bernert A, Chatterji T, Thalmeier P and Fulde P, 2001 *Eur. Phys. J.* B **21** 535.

[132] Lakkis S, Schlenker C, Chakraverty B K, Buder R and Marezio M, 1976 *Phys. Rev.* B **14** 1429.

[133] Fritsch V, Hemberger J, Brando M, Engelmayer A, Horn S, Klemm M, Knebel G, Lichtenberg F, Mandal P, Mayr F, Nicklas M and Loidl A, 2001 *Phys. Rev.* B **64** 045113.
Keimer B, Casa D, Ivanov A, Lynn J W, von Zimmermann M, Hill J P, Gibbs D, Taguchi Y and Tokura Y, 2000 *Phys. Rev. Lett.* **85** 3946.

[134] Abrahams E and Varma C M, *arXiv:cond-mat*/0003135.

[135] Hussey N E, Nakamae S, Behnia K, Takagi H and Urano C, 2000 *Phys. Rev. Lett.* **85** 4140.

[136] Ando Y, Boebinger G S, Passner A, Kimura T and Kishio K, 1995 *Phys. Rev. Lett.* **75** 4662. (LSCO)
Ando Y, Boebinger G S, Passner A, Wang N L, Greibel C, Steglich F, 1996 *Phys. Rev. Lett.* **77** 2065. (BSCCO)

[137] see Suter A, Mali M, Roos J and Brinkmann D, 2000 *Phys. Rev. Lett.* **84** 4938, and refs. therein.

[138] Meingast C, Karpinski J, Jilek E and Kaldis E, 1993 *Physica* C **209** 591.

[139] Sendyka T R, Dmowski W, Egami T, Seiji N, Yamauchi H and Tanaka S, 1995 *Phys. Rev. Lett.* **51** 6747.

[140] Wilson J A, 1977 *Structure and Bonding* **32** 57-91.

[141] Cooley J C, Aronson M C, Fisk Z and Canfield P C, 1995 *Phys. Rev. Lett.* **74** 1629.

[142] Travaglini G and Wachter P, 1984 *Phys. Rev.* B **30** 5877.

[143] Puchkov A V, Timusk T, Karlow M A, Cooper S L, Han P D and Payne D A, 1996 *Phys. Rev.* B **54** 6686.

[144] Degiorgi L, Nicol E J, Klein O, Grüner G, Wachter P, Huang S-M, Wiley J and Kaner R B, 1994 *Phys. Rev.* B **49** 7012.

[145] Lee Y S, Lee J S, Kim K W, Noh T W, Yu J, Choi E J, Cao G and Crow J E, 2001 *Europhys. Lett.* **55** 280.

[146] Pan S H, O'Neal J P, Badzey R L, Chamon C, Ding H, Engelbrecht J R, Wang Z, Eisaki H, Uchida S, Gupta A K, Ng K-W, Hudson E W, Lang K M and Davis J C, 2001 *arXiv:cond-mat*/0107347 {STM/STS}

[147] Yoshimura K, Imai T, Shimizu T, Ueda Y, Kosuge K and Yasuoka H, 1989 *J. Phys. Soc. Japan* **58** 3057. {NQR}

[148] Ryder J, Midgley P A, Exley R, Beynon R J, Yates D L, Afalfiz L and Wilson J A, 1991 *Physica* C **173** 9-24.

[149] Brinckmann J and Lee P A, 2001 *arXiv:cond-mat*/0107138.
Abanov A, Chubukov A V and Schmalian J, 2001 *Europhys. Lett.* **55** 369.

Abanov A, Chubukov A V and Schmalian J, 2001 *arXiv:cond-mat*/0107421.

[150] Chakravarty S and Kivelson S A, 2001 *Phys. Rev.* B **64** 064511.

[151] Gyorffy B L, Szotek Z, Temmerman W M, Andersen O K and Jepsen O, 1998 *Phys. Rev.* B **58** 1025.

Temmerman W M, Szotek Z, Gyorffy B L, Andersen O K and Jepsen O, 1996 *Phys. Rev. Lett.* **76** 307.

Suvasini M B, Temmerman W M and Gyorffy B L, 1993 *Phys. Rev.* B **48** 1202.

[152] Andersen O K, Liechtenstein A I, Jepson O and Paulsen F, 1995 *J. Phys. Chem. Solids* **56** 1573.

Anderson O K, Jepsen O, Liechtenstein A I and Mazin I I, 1994 *Phys. Rev.* B **49** 4145.

[153] Szotek Z, Gyorffy B L, Temmerman W M and Andersen O K, 1998 *Phys. Rev.* B **58** 522.

Szotek Z, Gyorffy B L and Temmerman W M, 2000 *Phys. Rev.* B **62** 3997.

Szotek Z, Gyorffy B L and Temmerman W M, 2001 *Physica* C **353** 23.

[154] Zheng-Johansson J and Wilson J A, 2001 *J. Phys.: Condens. Matter*, to be published.

[155] Gofron K, Campuzano J C, Abrikosov A A, Lindros M, Bansil A, Ding H, Koelling D and Dabrowski B, 1994 *Phys. Rev. Lett.* **73** 3302.

King D M, Zhen Z-X, Dessau D S, Marshall D S, Pack C H, Spicer W E, Peng J L, Li Z Y and Greene R L, 1994 *Phys. Rev. Lett.* **73** 3298.

[156] Newns D M, Tsuei C C, Pattnaik P C and Kane C L, 1992 *Comments Cond. Mat. Phys.* **15** 263-272. (Gordon & Breach).

[157] Khodel V A, Shaginyan V R and Shak P, 1996 *JETP Lett.* **63** 752.

[158] Wijngaarden R J, Tristan Jover D and Griessen R, 1999 *Physica* B **265** 128.

[159] Temmerman W M, Winter H, Szotek Z and Svane A, 2001 *Phys. Rev. Lett.* **86** 2435.

[160] Burrow J H, Maule C H, Strange P, Tothill J N and Wilson J A, 1987 *J. Phys: Solid State Phys.* **20** 4115-4133.

[161] Anisimov V I, Elfimov I S, Hamada N and Terakura K, 1996 *Phys. Rev.* B **54** 4387.

García J, Subías G, Proietti M G, Blasco J, Renevier H, Hodeau J L and Joly Y, 2001 *Phys. Rev.* B **63** 054110.

[162] Menushenkov A P and Klementev K V, 2000 *J. Phys.: Condens. Matter* **12** 3767.

Braden M, Reichardt W, Elkaim E, Lauriat J P, Shiryaev S and Barilo S N, 2000 *Phys. Rev.* B **62** 6708.

See Winkler B, Pickard C J, Segall M D and Milman V, 2001 *Phys. Rev.* B **63** 214103, for the comparable $Au_IAu_{III}$ charge ordering in perovskite $CsAuCl_3$.

[163] Pearson W B, '*Crystal Chemistry of Metals and Alloys'.* (Pub: Wiley 1972).

[164] Thiele G, Steiert M, Wagner D and Wochner H, 1984 *Zeit. Anorg. (Allg.) Chem.* **516** 207.

[165] Edwards H L, Barr A L, Markert J T and de Lozanne A L, 1994 *Phys. Rev. Lett.* **73** 1154.

[166] Dahl J.P and Switendick A C, 1966 *J. Phys. Chem. Sol.* **27** 931. {$Cu_2O$}

Kleinman L and Mednick K, 1980 *Phys. Rev.* B **21** 4549; {$Cu_2O$}
1979 *Phys. Rev.* B **20** 2487. {CuCl}

[167] Zhang S B, Wei S-H and Zunger A, 1995 *Phys. Rev.* B **52** 13975.
Rohlfing M, Kruger P and Pollman J, 1998 *Phys. Rev.* B **57** 6485.

[168] Lambrecht W R L, Segall B, Strite S and Martin G, 1994 *Phys. Rev.* B **50** 14155.

[169] Forsyth J B, Brown P J and Wanklyn B M, 1988 *J. Phys. C: Solid State Phys*. **21** 2917.

[170] Hudson E W, Lang K M, Madhavan V, Pan S H, Eisaki H, Uchida S and Davis J C, 2001 (May) preprint.
Pan S H *et al,* 2000 *Nature* **403** 746.

[171] Singer P M, Hunt A W and Imai T, 2001 *arXiv:cond-mat*/0108291.
Haase J, Slichter C P, Stern P, Milling C T and Hinks D G, 2000 *J. Supercond.* **13** 723.

[172] Hussey N E, 2001 (Aug.) preprint.

[173] Kim M S, Choi J H and Lee S I, 2001 *Phys. Rev.* B **63** 092507.

[174] Feng D L, Kim C, Eisaki H, Lu D H, Shen K M, Ronning F, Armitage N P, Damascelli A, Kaneko N, Greven M, Shimoyama J-i, Kishio K, Yoshizaki R, Gu G D and Shen Z-X, 2001 *arXiv:cond-mat*/0107073. [2212]
but note - Janowitz C, Müller R, Dudy L, Krapf A, Manzke R, Ast C and Höchst H, 2001 *arXiv:cond-mat*/0107089. [2201]

[175] Grevin B, Berthier Y, Collin G and Mendels P, 1998 *Phys. Rev. Lett.* **80** 2405.

[176] Wakimoto S, Birgeneau R J, Lee Y S and Shirane G, 2001 *Phys. Rev.* B **63** 172501.

[177] Mook H A, Dai P and Dogan F, 2001 *Phys. Rev.* B **64** 012502.

[178] Varma C M, 1999 *Phys. Rev. Lett.* **83** 3538
Chakravarty S, Laughlin R B, Morr D K and Nayak C, 2001 *Phys. Rev.* B **64** 094503.

[179] Sonier J E, Brewer J H, Kiefl R F, Haffner R H, Poon K, Stubbs S L, Morris G D, Miller R I, Hardy W N, Liang R, Bonn D A, Gardner J S and Curro N J, 2001 *arXiv:cond-mat*/0108479.

[180] Moessner R, Sondhi S L and Fradkin E, *arXiv:cond-mat*/0103396.

[181] Gyorffy B L, Staunton J B and Stocks G M, 1991 *Phys. Rev.* B **44** 5190.
Litak G and Gyorffy B L, 2000 *Phys. Rev.* B **62** 6629.

[182] Chen Z B, Levin K and Kosztin I, 2001 *Phys. Rev.* B **63** 184519 & refs. therein.

[183] Homes C C, Kamal S, Bonn D A, Liang R, Hardy W N and Clayman B P, 1998 *Physica* C **296** 230.

[184] Micnas R, Ranninger J and Robaszkiewicz S, 1990 *Rev. Mod. Phys.* **62** 113.

[185] Domanski T and Ranninger J, 2001 *Phys. Rev*. B **63** 134505.
Romano A, 2001 *arXiv:cond-mat*/0106170

[186] Friedberg R and Lee T D, 1989 *Phys. Rev.* B **40** 6745.

[187] Ren H-C, 1998 *Physica* C **303** 115.
Blaer A S, Ren H C and Tchernyshyov O, 1997 *Phys. Rev.* B **55** 6035.
Piegari E, Cataudella V and Iadonisi G, 1998 *Physica* C **303** 273.

[188] Casas M, Davidson N J, de Llano M, Mamedov T A, Puente A, Quick R M, Rigo A and Solis M A, *arXiv:cond-mat*/0102243.

Fortes M, Solis M A, de Llano M and Tolmachev V V, *arXiv:cond-mat*/0103398.